\documentclass[aps,prc,preprintnumbers,showpacs,superscriptaddress,amsmath,amssymb,showkeys,nofootinbib]{revtex4}
\usepackage{graphicx}
\usepackage{dcolumn}
\usepackage{bm}
\usepackage{epsfig}
\usepackage{float}

\begin{document}

\title{Signatures of the chiral two-pion exchange electromagnetic currents in 
the $^2$H and $^3$He photodisintegration reactions}
\author{D.~Rozp\c{e}dzik}
\affiliation{M. Smoluchowski Institute of Physics, Jagiellonian University, PL-30059 Krak\'ow, Poland}
\author{J.~Golak}
\affiliation{M. Smoluchowski Institute of Physics, Jagiellonian University, PL-30059 Krak\'ow, Poland}
\author{S.~K\"olling}
\affiliation{Forschungszentrum J\"ulich, Institut f\"ur Kernphysik (IKP-3) and
J\"ulich Center for Hadron Physics, \\ D-52425 J\"ulich, Germany}
\affiliation{Helmholtz-Institut f\"ur Strahlen- und Kernphysik (Theorie)
and Bethe Center for Theoretical Physics,\\ Universit\"at Bonn, D-53115 Bonn, Germany}
\author{E.~Epelbaum}
\affiliation{Institut f\"ur Theoretische Physik II, Ruhr-Universit\"at Bochum, D-44780 Bochum, Germany}
\author{R.~Skibi\'nski}
\affiliation{M. Smoluchowski Institute of Physics, Jagiellonian University, PL-30059 Krak\'ow, Poland}
\author{H.~Wita{\l}a}
\affiliation{M. Smoluchowski Institute of Physics, Jagiellonian University, PL-30059 Krak\'ow, Poland}
\author{H.~Krebs}
\affiliation{Institut f\"ur Theoretische Physik II, Ruhr-Universit\"at Bochum, D-44780 Bochum, Germany}

\begin{abstract}
The recently derived long-range two-pion exchange
(TPE) contributions to the nuclear current operator which appear at next-to-leading order (NLO) of the
chiral expansion are used to describe electromagnetic processes.
We study their role in the photodisintegration of $^2$H and $^3$He and compare our
predictions with experimental data. The bound and scattering states
are calculated using five different parametrizations of the chiral next-to-next-to-leading order (N$^2$LO) nucleon-nucleon (NN) potential
which allows us to estimate the theoretical uncertainty at a given order in the
chiral expansion. For some observables the results are very close to the
predictions based on the AV18 NN potential and the current operator
(partly) consistent with this force. In the most cases, the addition of long-range TPE currents improved the description of the experimental data.
\end{abstract}

\date{\today}

\maketitle
\section{Introduction}
\def\theequation{\arabic{section}.\arabic{equation}}
\setcounter{equation}{0}

Chiral Effective Field Theory (ChEFT) provides a systematic and 
model-independent framework to analyze 
hadron structure and dynamics in harmony with the spontaneously 
broken approximate chiral symmetry of QCD. This approach is a powerful 
tool for the derivation of the nuclear forces. Exchange vector and axial 
currents in nuclei have also been studied in the framework of ChEFT. 
Since the pioneering work of Park et al.~\cite{park:1993}, heavy-baryon chiral 
perturbation theory has been applied to derive exchange axial and vector 
currents for small values of the photon momentum. These calculations have been
carried out in time-ordered perturbation theory. The resulting exchange vector 
currents have been, in particular, applied to analyze 
radiative neutron-proton capture within a hybrid approach~\cite{park:2000}.

ChEFT has also been used to study the electromagnetic properties
of the deuteron~\cite{Walzl:2001vb,Philips:2009} and $^3$He~\cite{romek1,shukla:2009}.
One of the fundamental processes observed for the deuteron is the photodisintegration reaction. 
It has been a subject of intense experimental and theoretical research for
several decades (see Refs~\cite{arenhovel:1991, gilman:2002}).
Also photodisintegration of $^3$He has been studied experimentally and
theoretically for a long
time~\cite{carlson:1998,marccuci:2009,golak:2005}. Photodisintegration
observables provide a good tool for studying 
the contributions from meson exchange currents (MEC) to the nuclear current operator.
This is because the charge density operator, which often dominates low-energy electrodisintegration
and is mostly given by the single nucleon current, does not play any role in this reaction.
An ongoing interest in low-energy photodisintegration reactions, especially in
view of planned experiments, provides a strong motivation to apply the
framework of chiral effective field theory.  This approach relies on the approximate spontaneously
broken chiral symmetry of QCD. It allows for a systematic derivation of the nuclear Hamiltonian
and the corresponding electromagnetic current operator from the underlying effective
Lagrangian for pions and nucleons via the chiral expansion, i.e. a
simultaneous expansion in soft momenta of external particles and about the
chiral limit. For more details on the application of ChEFT to nuclear forces
and currents the reader is referred to recent review articles \cite{evgeny:2006,Epelbaum:2008ga} and references therein.  

In the two- and three-nucleon systems, the leading contributions 
to the exchange current originate from one-pion exchange which are well known. 
The 2N current operator at the leading loop order in the chiral expansion 
has been worked out by Pastore et al.~\cite{pastore:2008,Pastore:2009is} based on time-ordered perturbation
theory. Independently, the two-pion exchange 2N current operator has been derived
in Ref.~\cite{TPE:2009} using the method of unitary transformation. The resulting current operator is consistent with the corresponding chiral
two-nucleon potential~\cite{evgeny:2006} obtained within the same scheme. In the present work, for the first time 
we explore the effects of the leading two-pion exchange 2N 
operator~\cite{TPE:2009} in the photodisintegration reactions of $^2$H and $^3$He. We, however, emphasize that
the presented calculations are not yet complete. In particular, the
corresponding expressions for the one-pion exchange at NLO and short-range contributions to the
current operator within the method of unitary transformation are not yet
available. Our main goal in the present work is to explore the sensitivity of
various observables in the deuteron and $^3$He photodisintegration to the two-pion
exchange current rather than to provide a complete description of these
reactions within the ChEFT framework. 

Our manuscript is organized as follows.  
In section II the formalism which we use to describe selected 2N electromagnetic reactions is presented.
The results for the photodisintegration of the $^2$H are 
discussed and compared with the experimental data 
in section III. The extension to the 3N system is briefly described in section IV and the results obtained 
for photodisintegration of $^3$He are presented in Section V.  
Finally, section VI contains the summary and conclusions.

\section{Formalism}
\def\theequation{\arabic{section}.\arabic{equation}}
\setcounter{equation}{0}
The general form of the nuclear matrix element for electromagnetic disintegration reactions in the 2N system is represented by
\begin{equation}
\label{eq3}
 N^{\mu}\equiv \langle \Psi^{2N}_{\rm scatt}|J^{\mu}(\vec Q)|\Psi^{2N}_{\rm bound}\rangle \,,
\end{equation}
where  the proton-neutron scattering state $|\Psi^{2N}_{\rm scatt}\rangle$ and 
the deuteron bound state $|\Psi^{2N}_{\rm bound}\rangle$ are obtained using NN potential. 
The current operator $J^{\mu}(\vec Q)$ acts between the internal initial and final 2N states.
We employ the solution of the Lippmann-Schwinger equation,
$t = V_{2N} + tG_0 V_{2N}$, in order to express $N^{\mu}$ as
\begin{equation}
\label{eq5}
 N^{\mu} = \langle \vec{p_{0}} \mid \left( 1 + t G_0 \right) \, J^{\mu}(\vec
 Q)\mid \Psi^{2N}_{\rm bound}\rangle ,
\end{equation}
where $G_{0}$ is the free 2N propagator, $t$ is the NN t-matrix 
and $|\vec{p}_{0}\rangle$ is the eigenstate of the relative proton-neutron momentum.
Since all observables can be computed from $N^{\mu}$, the description of the electromagnetic reactions requires the 
knowledge of the consistent potential and electromagnetic current.
The NN potential based on ChEFT is currently available up 
to next-to-next-to-next-to-leading order in the chiral expansion~\cite{evgeny:2006,Epelbaum:2008ga}.
As already pointed out, in this paper we focus on the long-range TPE contributions to the current operator, which appear at NLO.
However, in order to avoid the theoretical error from using the less accurate NLO NN potential, 
all calculations are made using the N$^2$LO potential.
At this order, the NN potential $V_{2N}$ is built from the one-pion exchange (OPE), $V_{1\pi}$, and two-pion 
 exchange (TPE), $V_{2\pi}$, contributions as well as various contact
interactions ($\rm cont$)~\cite{evgeny:2006}
\begin{equation}
 V_{2N} = V_{1\pi} + V_{2\pi} + V_{\rm cont} \,.
\end{equation}
The effective current operator $J^\mu$ for the 2N system is a sum of the 
single-nucleon operators $ J^{\mu}(i),i=1,2$ and two-nucleon operators
of different type ($J^{\mu}(1, 2)$)
\begin{eqnarray}
 J^{\mu} = J^{\mu}(1) + J^{\mu}(2) + J^{\mu}(1, 2)\,,
\end{eqnarray}
where
\begin{equation}
J^{\mu}(1, 2) = J^{\mu}_{1\pi}(1, 2) + J^{\mu}_{2\pi}(1, 2) +  J^{\mu}_{\rm cont}(1, 2).
\end{equation}
The expressions for the single-nucleon and the leading OPE 
currents $J^{\mu}_{1\pi}(1, 2)$ are well established, see e.g.~\cite{ope}. The
results for the leading two-pion exchange contributions used in the present
work are available in Refs.~\cite{pastore:2008,TPE:2009}. We emphasize that the resulting two-pion exchange
current is parameter-free. 
The expressions for the OPE and contact currents at the leading loop level have
been recently worked out within time-ordered perturbation theory \cite{Pastore:2009is}.  
Work on the derivation of these contributions using the method of unitary
transformation  is still in progress.

The 2N four-current operator $J^{\mu}(1, 2)\equiv(J^{0}(1, 2),\vec{J}(1, 2))$ can be decomposed according to its
isospin and spin-momentum structure and quite generally written in the form~\cite{TPE:2009,foldy:1979}
\begin{eqnarray}
\label{eq2}
J^{0} (1, 2) &=& \sum_{\eta=1}^{5} \sum_{\beta=1}^{8}
f_{\eta}^{\beta S}(\vec{q}_1,\vec{q}_2) 
T_{\eta} O^{S}_{\beta}\,,\\ 
\label{eq2a}
\vec{J} (1, 2) &=& \sum_{\eta=1}^{5} \sum_{\beta=1}^{24}
f_{\eta}^{\beta}(\vec{q}_1,\vec{q}_2) 
T_{\eta} \vec{O}_{\beta}\,, 
\end{eqnarray}
where $\vec{q}_{i}\equiv \vec{p}\,'_i-\vec{p_i}$ is the momentum transferred to nucleon $i$, 
$T_{\eta}$ is the 2N isospin operator, $O^S_{\beta}$ and $ \vec{O}_{\beta} $ 
are the (momentum dependent) spin operators in the 2N space, $f_{\eta}^{\beta S}$ 
and $f_{\eta}^{\beta}$ are scalar functions. The explicit form of the scalar
functions and the operator basis for $O_\beta^S$ and $\vec O_\beta$ can be found in Ref.~\cite{TPE:2009}.

In this paper, we concentrate on a treatment of the long-range TPE
contributions to the 2N current operator derived in Ref.~\cite{TPE:2009}.
The expressions for the functions $f_{\eta}^{\beta S}(\vec{q}_1,\vec{q}_2)$
and $f_{\eta}^{\beta}(\vec{q}_1,\vec{q}_2)$ entering the TPE current and
charge density operators in Eqs.~(\ref{eq2}), (\ref{eq2a}) are rather
complicated and contain the standard loop
functions and the three-point functions in a form suitable for numerical 
calculations~\cite{TPE:2009}. 
Due to their isospin structure, not all combinations of (\ref{eq2}) and (\ref{eq2a}) contribute to photodisintegration 
of the deuteron. The non-vanishing contributions emerge from 
\begin{equation}
\label{J2PI2N}
 \vec{J}_{2\pi}(1, 2) = \sum_{\beta=3}^{10}
f_{2}^{\beta}(\vec{q}_1,\vec{q}_2) \, 
(\vec \tau_1 - \vec \tau_2)_3  \, \vec{O}_{\beta} + f_{3}^{2}(\vec{q}_1,\vec{q}_2)
\, ( \vec \tau_1 \times \vec \tau_2 )_3 \,  \vec{O}_{2}\,,
\end{equation}
where $( ... )_3$ denotes the third cartesian component of the vector.
We work in momentum space and apply the standard partial wave decomposition
of the 2N potential, see e.g.~Ref.~\cite{gloekle:1983} for more details. 
Our calculations are performed using a complete
set of 2N states
\begin{equation}
 |p\alpha\rangle\equiv|p(ls)jm_j\rangle|tm_t\rangle
\end{equation}
where $p$ is the magnitude of the relative momentum, $l$, $s$, $j$ and $m_j$ are the orbital angular momentum, 
spin, total angular momentum and its projection on the quantisation axis $\hat z$, respectively.
The isospin quantum numbers of the two-nucleon system are denoted by $t$ and $m_t$. 

The TPE current operator needs to be expressed in the same partial wave basis. To this end,
we first prepare all spin and isospin matrix elements using $\textit{Mathematica}\copyright$
and then calculate the resulting four-fold angular integrals
\begin{eqnarray}
\label{matrix}
 \langle \,p'\alpha'| \vec J_{2\pi}(1, 2)|p\alpha \rangle &=& \langle p'(l's')j'm_{j'};t'm_{t'} | {\vec J}_{\eta\beta} | p(ls)jm_j;tm_{t}\rangle \\ \nonumber
&=& \int\nolimits {\rm d} \hat p' \, {\rm d} \hat p \sum_{m_l,m_l'}
C(l's'j';m_{l'},m_{j'}-m_{l'},m_{j'})  
\,  Y_{l'm_{l'}}^{*}(\hat p') \, C(lsj;m_{l},m_j-m_{l},m_j) \, Y_{lm_{l}}(\hat
p) \\ \nonumber 
&\times&f^{\beta}_{\eta}(\vec{q_{1}},\vec{q_{2}})\, 
\langle t'm_{t'}|T_{\eta}|tm_{t}\rangle \, \langle s'\,m_{j'}-m_{l'}|\vec O_{\beta}|s\,m_{j}-m_{l}\rangle \,,
\end{eqnarray}
numerically. Here, 
$C(lsj;m_{l},m_j-m_{l},m_j)$ denote the Clebsch-Gordon coefficient and $Y_{lm_{l}}(\hat p)$ are 
the spherical harmonics. Such an approach has been described in Ref.~\cite{golak:2010}. 
In order to calculate the four-fold integrals in Eq.~(\ref{matrix}) for 
the whole grids of $p$ and $p'$ points and all non-vanishing $(\alpha, \alpha', m_j)$ combinations 
we used the parallel supercomputer IBM Blue Gene/P of the J\"ulich Supercomputing Centre (JSC).

\section{Results for photodisintegration of the deuteron}
\begin{figure}[!ht]
     \begin{tabular}{lcr}
\hspace{0mm}\includegraphics[width=6.1cm,height=5.3cm]{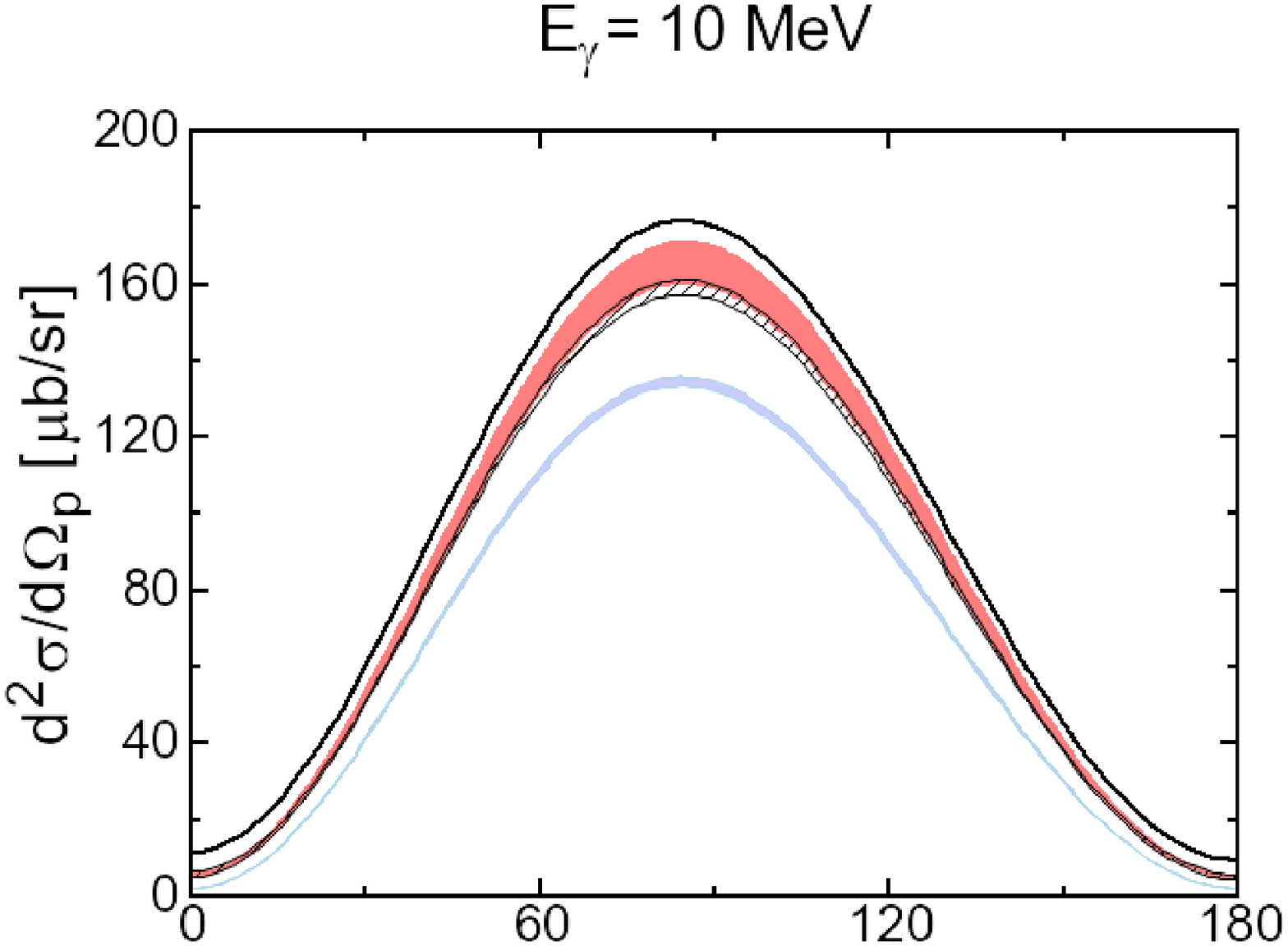} &
\hspace{0mm}\includegraphics[width=5.7cm,height=5.3cm]{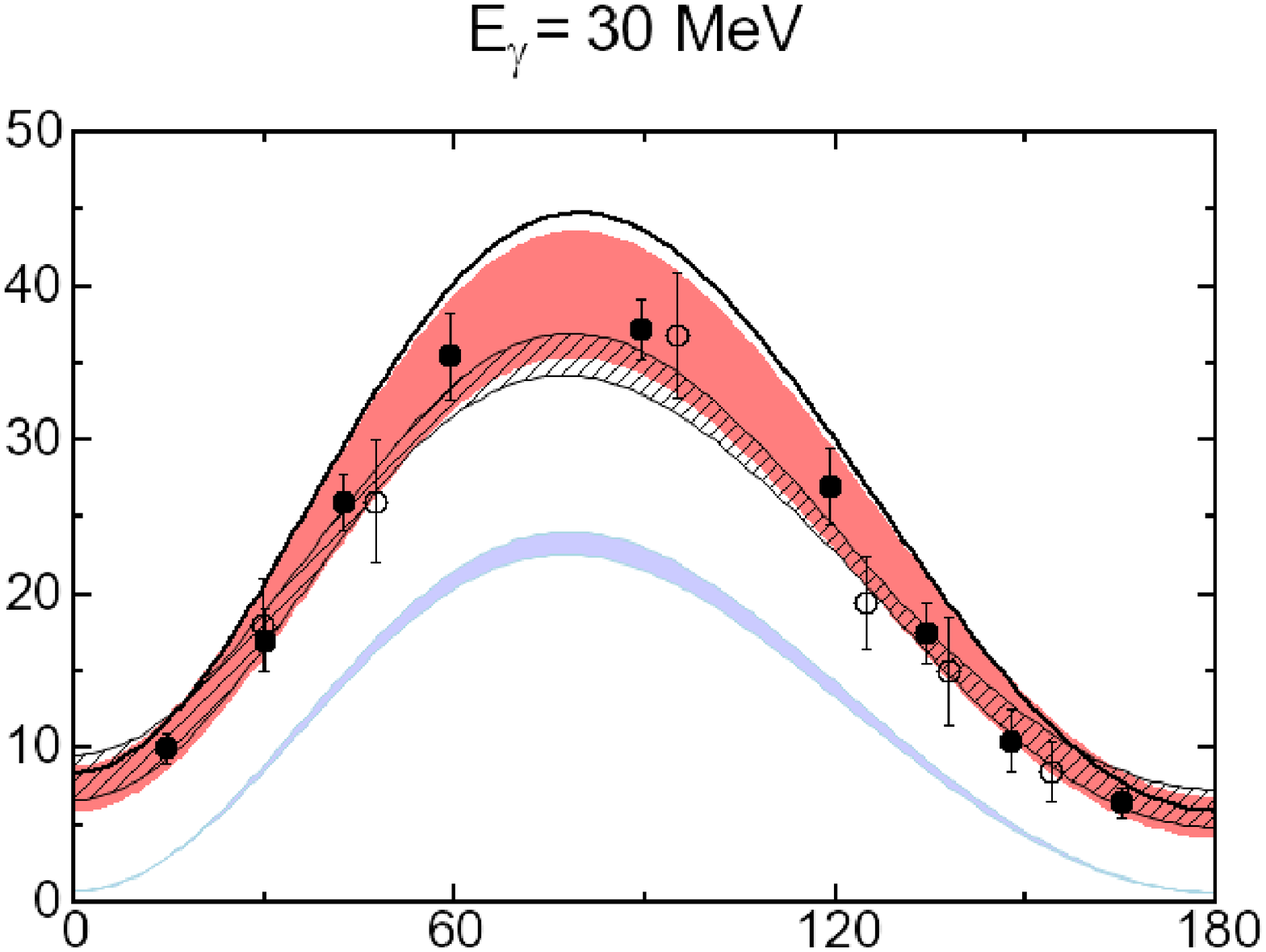} &
\hspace{0mm}\includegraphics[width=5.7cm,height=5.3cm]{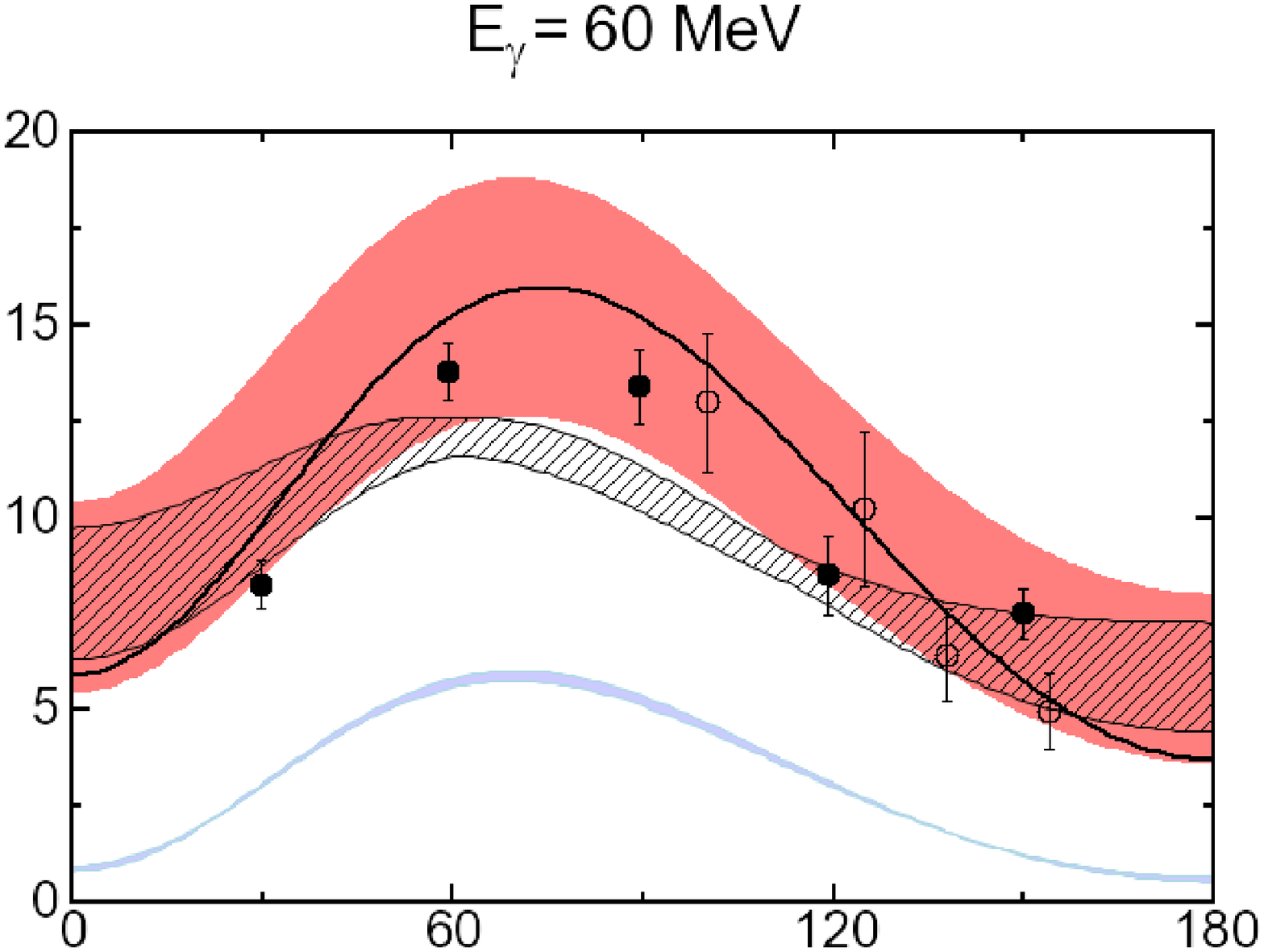}\\
\hspace{0mm}\includegraphics[width=6.1cm,height=4.8cm]{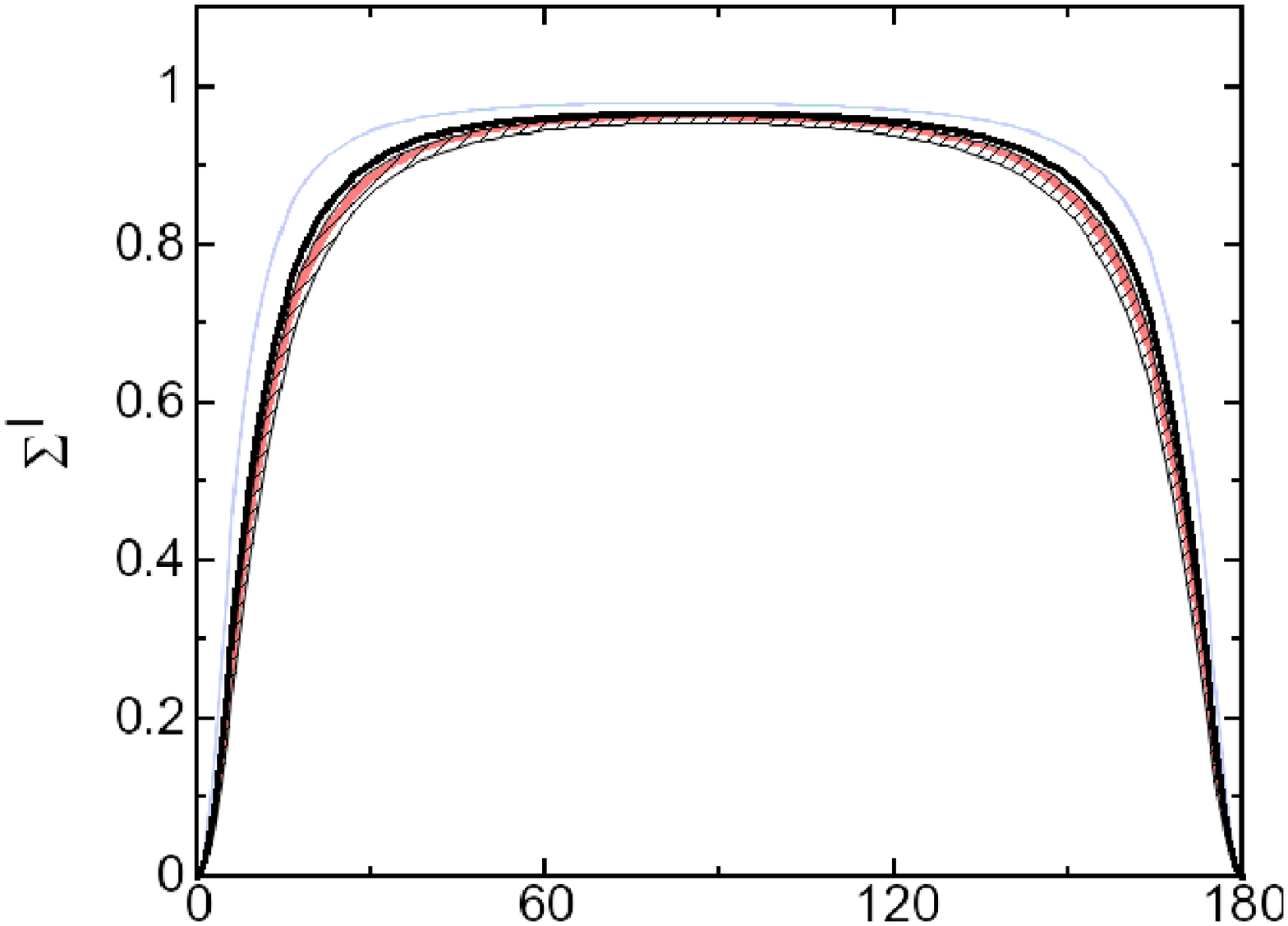} &
\hspace{0mm}\includegraphics[width=5.7cm,height=4.8cm]{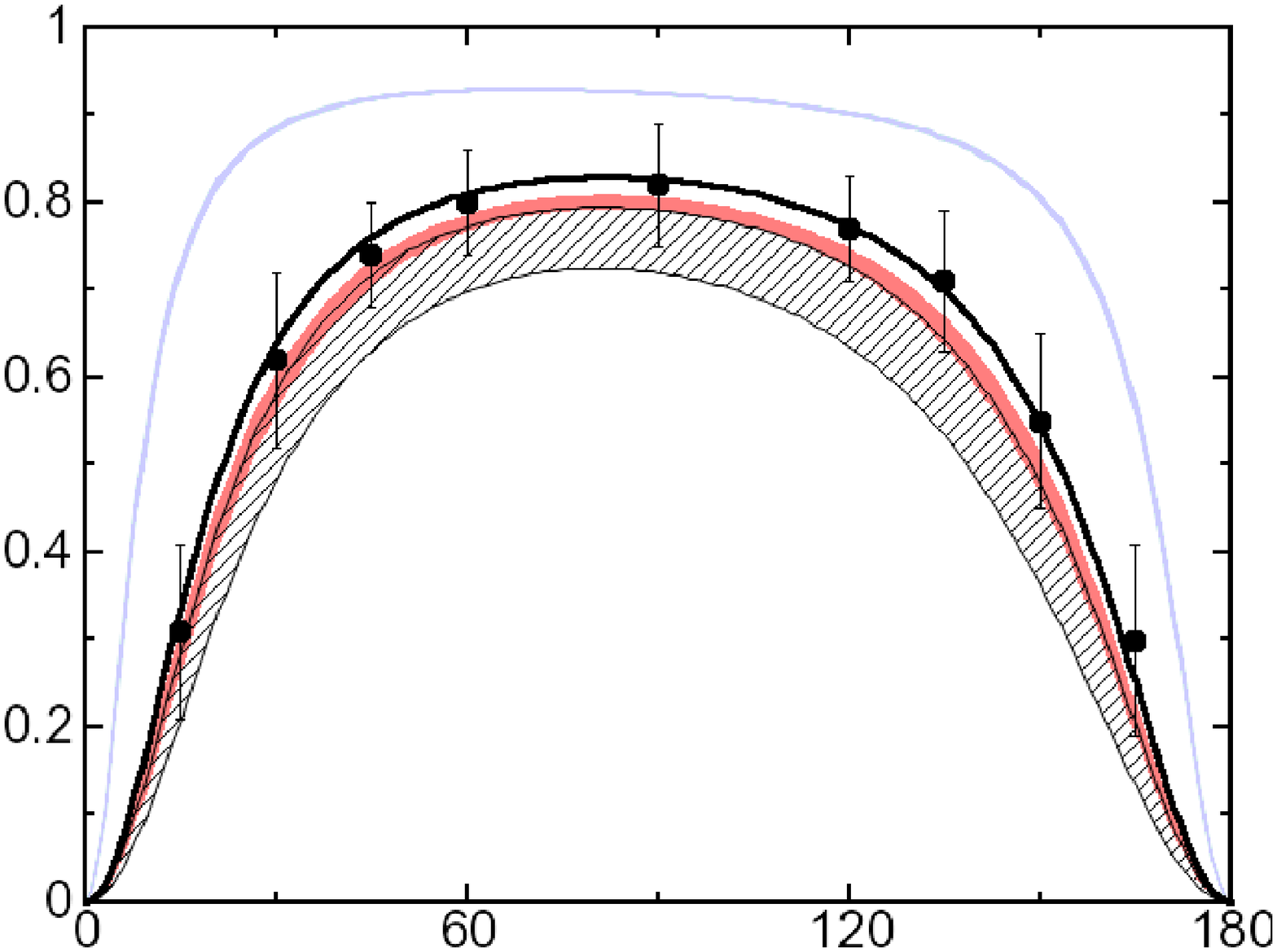} &
\hspace{0mm}\includegraphics[width=5.7cm,height=4.8cm]{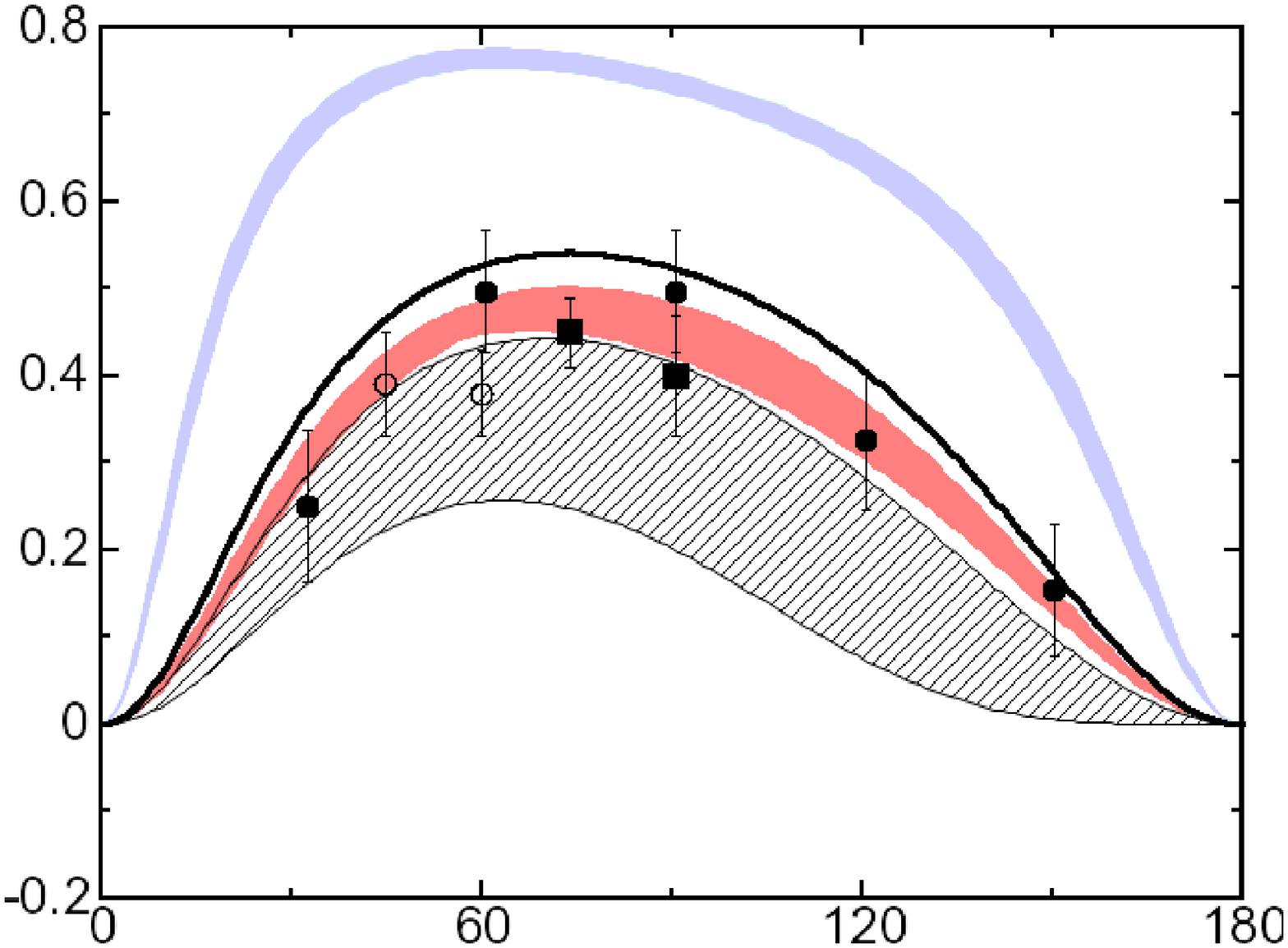} \\
\hspace{0mm}\includegraphics[width=6.1cm,height=5.1cm]{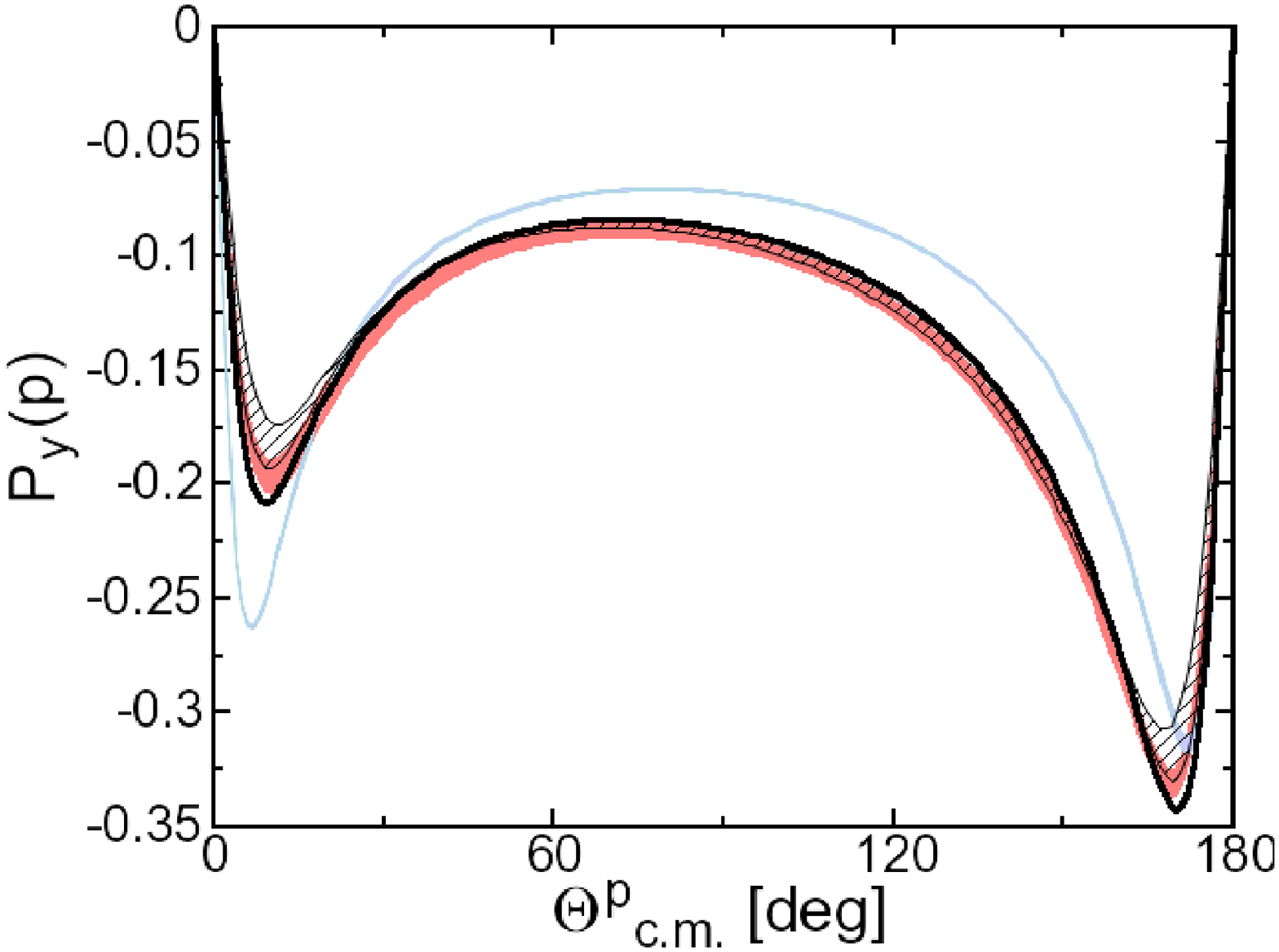} &
\hspace{0mm}\includegraphics[width=5.7cm,height=5.1cm]{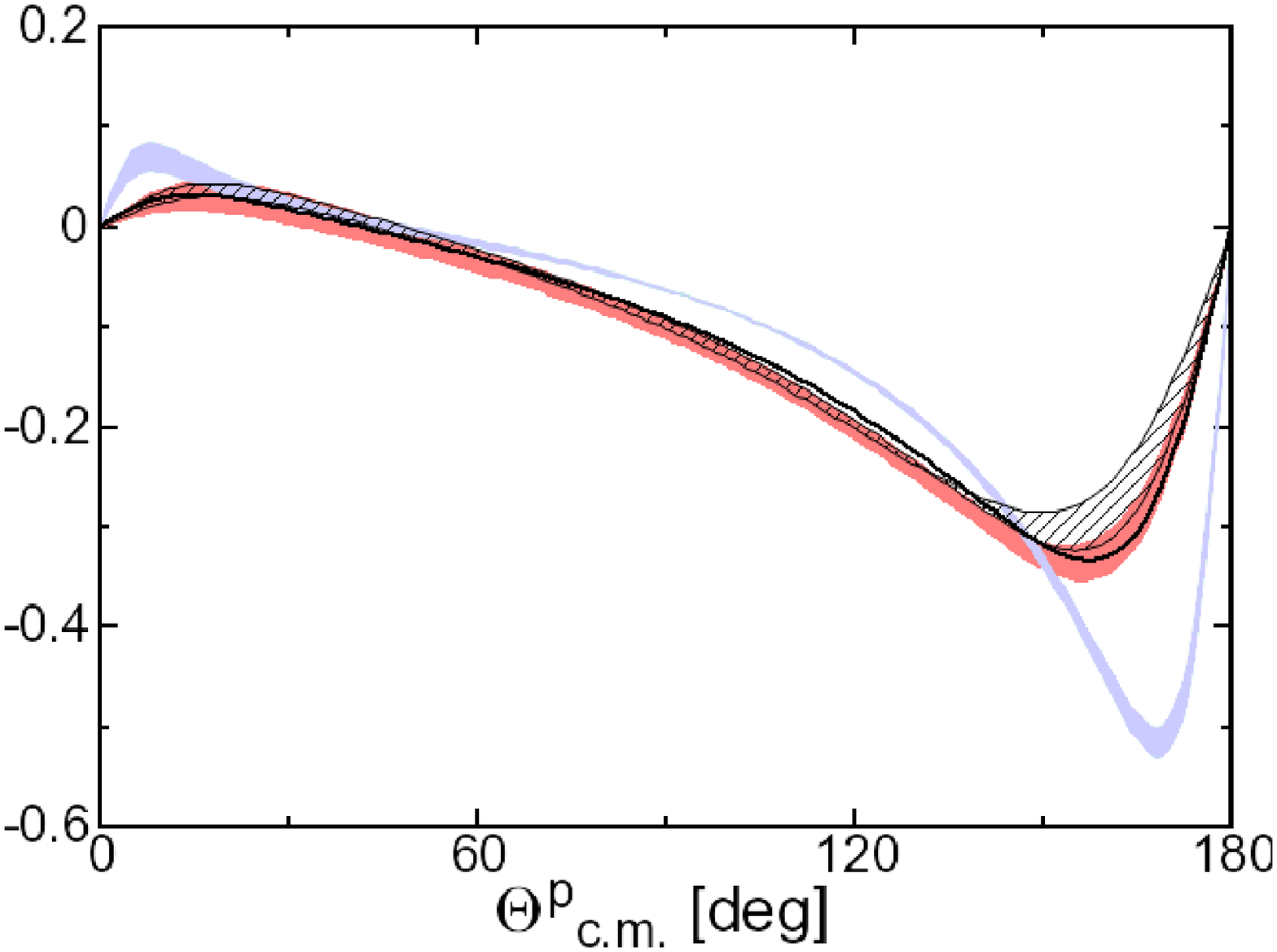} &
\hspace{0mm}\includegraphics[width=5.7cm,height=5.1cm]{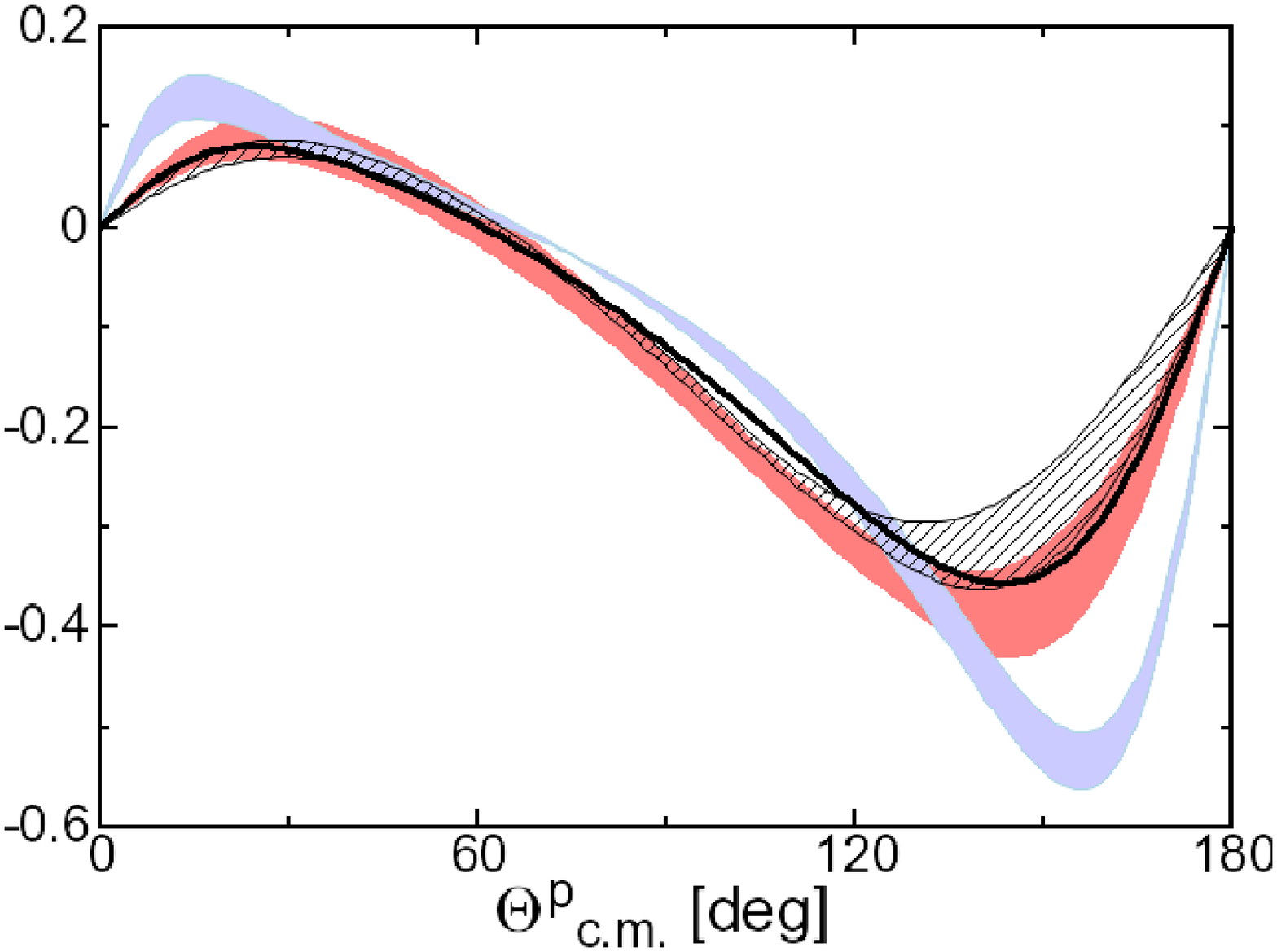} \\
   \end{tabular}
\caption{\label{fig1}(color online)
The results for the unpolarized cross section, the photon analyzing power and outgoing proton polarization in the
deuteron photodisintegration process at the photon laboratory energies of
$E_{\gamma}$= 10, 30, 60 MeV, displayed as functions of the proton emission
angle. The solid black line refers to  the standard calculation based on the
AV18 potential, the light (blue) band covers results obtained with the single-nucleon
current only, the hatched band represents the predictions based on the
single-nucleon and OPE parts and the dark (pink) band includes, in addition, the contributions
of the TPE current. The experimental data are from Ying et al.~\cite{ying}.}
\end{figure}

\def\theequation{\arabic{section}.\arabic{equation}}
\setcounter{equation}{0}
We now discuss the results for the deuteron photodisintegration process 
for the unpolarized cross section and selected polarization observables.
The results for the differential cross section, the photon analyzing power and outgoing proton polarization
at the photon laboratory energies of $E_{\gamma}$ = 10, 30 and 60 MeV are shown in Fig.~\ref{fig1}.
The bands reflect the uncertainty due to the variation of the two cut-off parameters
$\Lambda$ and $\tilde{\Lambda}$ that appear in the chiral potential.
While the first cut-off parameter $\Lambda$ appears in the regulator function for 
the Lippmann-Schwinger equation, 
the second parameter $\tilde{\Lambda}$ enters the spectral function regularization (SFR)
and denotes the ultraviolet cut-off value in the mass spectrum of the two-pion exchange potential. 
Following Ref.~\cite{evgeny:2006}, 
the cut-off values are varied between 450 and 600 MeV for $\Lambda$ and between 
500 and 700 MeV for $\tilde{\Lambda}$. It is important to emphasize that the
resulting bands probably overestimate the theoretical uncertainty that can be
expected in a complete calculation at this order in the chiral expansion. This
is because we have not yet included the corresponding short-range
contributions to the nuclear current, which are expected to absorb the large
part of the cut-off dependence.  Thus, the interpretation of the band width in
the obtained results in terms of the theoretical uncertainty should be taken
with care. Different bands shown in Fig.~\ref{fig1} describe the contributions
from the different parts of the 2N current: single-nucleon current (light
band), one-pion exchange contribution (hatched band) and the long-range TPE contributions (dark band).
As a reference, we also show the results based on the phenomenological AV18 potential~\cite{wiringa:1995}
and the corresponding current model~\cite{Riska1,Riska2}. 
Notice that the results obtained solely from the single nucleon currents and
by adding the OPE contributions do not describe the data well 
and differ significantly from the reference AV18 predictions.
An explicit inclusion of the TPE contributions yields an improved description
of the experimental data which turns out to be in agreement with the AV18 predictions. 
The OPE predictions give the cross section and photon analyzing power values lower than the AV18 results.
The bands including TPE currents are broad, however, for $E_\gamma=30$ and 60
MeV they give reasonable description of the experimental data. 
In the case of photon analyzing powers, the best agreement between all models is obtained at lowest energy. 
For energies $E_\gamma= 30$ and $60$ MeV,  calculations including the TPE
currents yield even better agreement with experimental data than the AV18
results. It remains to be seen whether this conclusion will still hold after 
including the short-range and the subleading OPE currents.  
It is further important to emphasize that the single nucleon current alone is
insufficient to describe the data
for this observable.  
Thus, it is necessary to include higher order electromagnetic currents.
In the case of the outgoing proton polarization, we observe a smaller
sensitivity to the TPE currents and a good agreement between traditional framework (AV18) 
and chiral results at all energies considered. The larger sensitivity to the
details of the exchange currents is only
observed at forward and backward outgoing proton angles.

\begin{figure}[!ht]
     \begin{tabular}{lcr}
     \hspace{-3mm}\includegraphics[width=6.0cm,height=4.7cm]{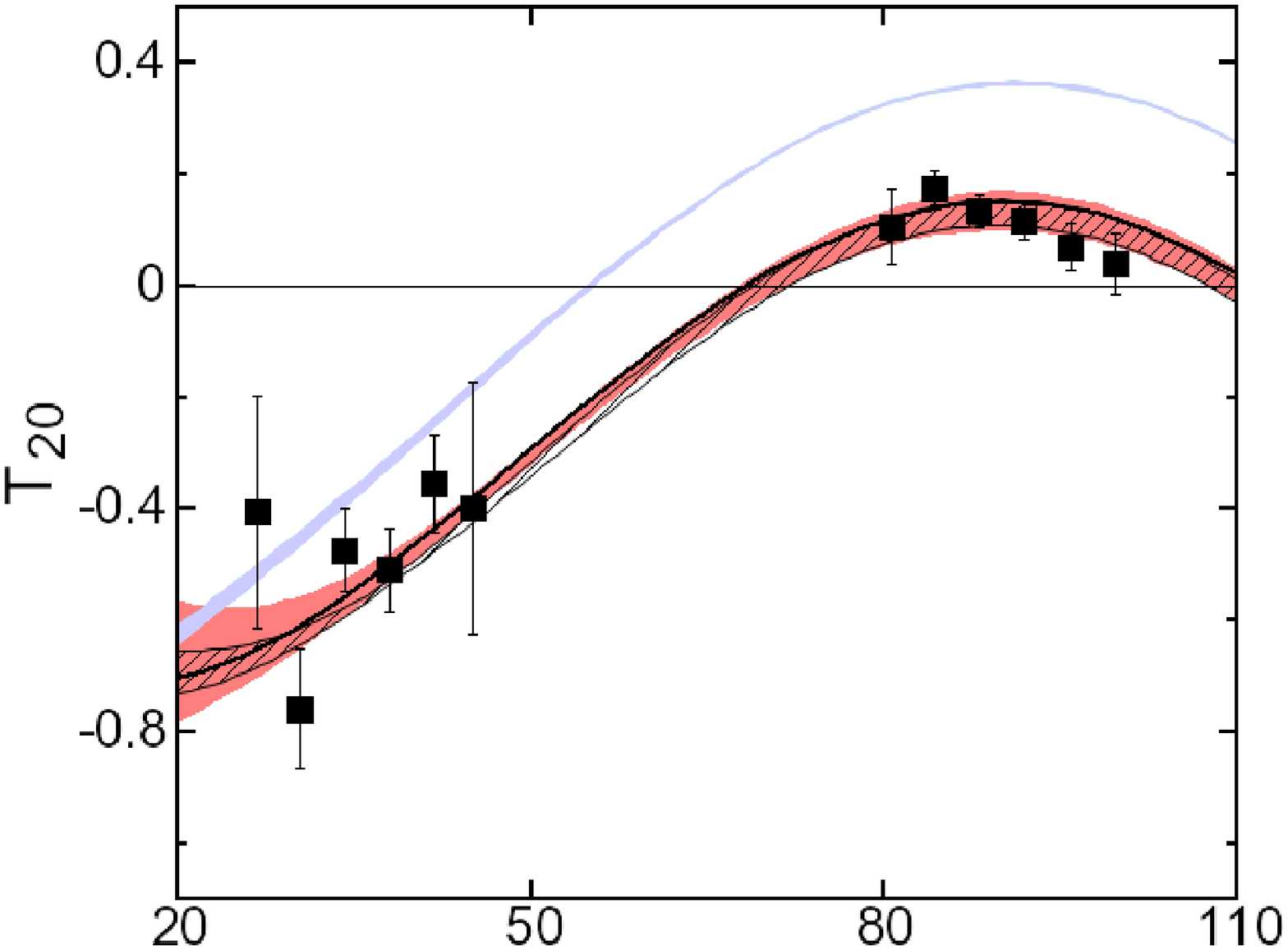} &
     \hspace{-1mm}\includegraphics[width=6.0cm,height=4.7cm]{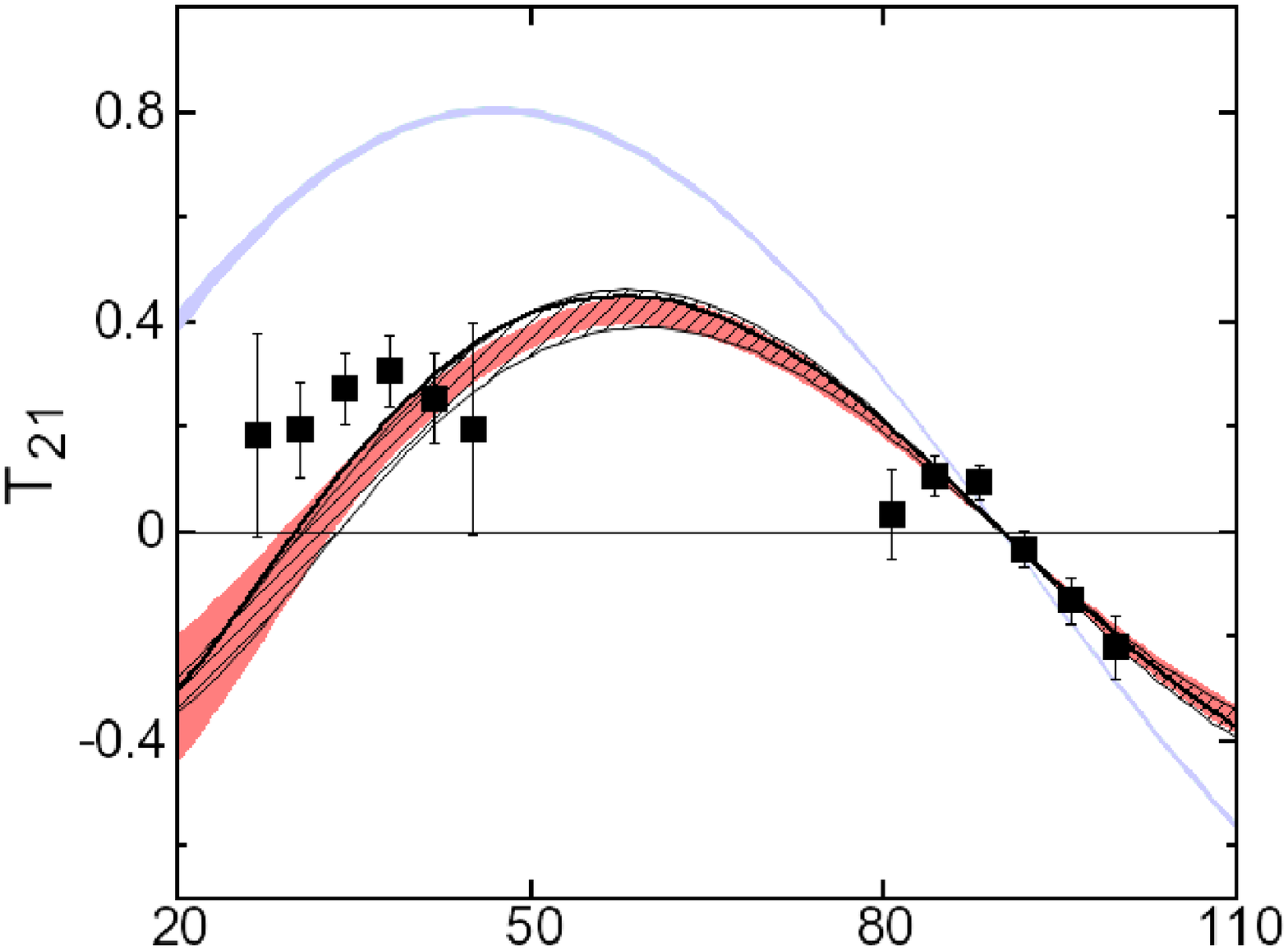} &
     \hspace{-1mm}\includegraphics[width=6.0cm,height=4.7cm]{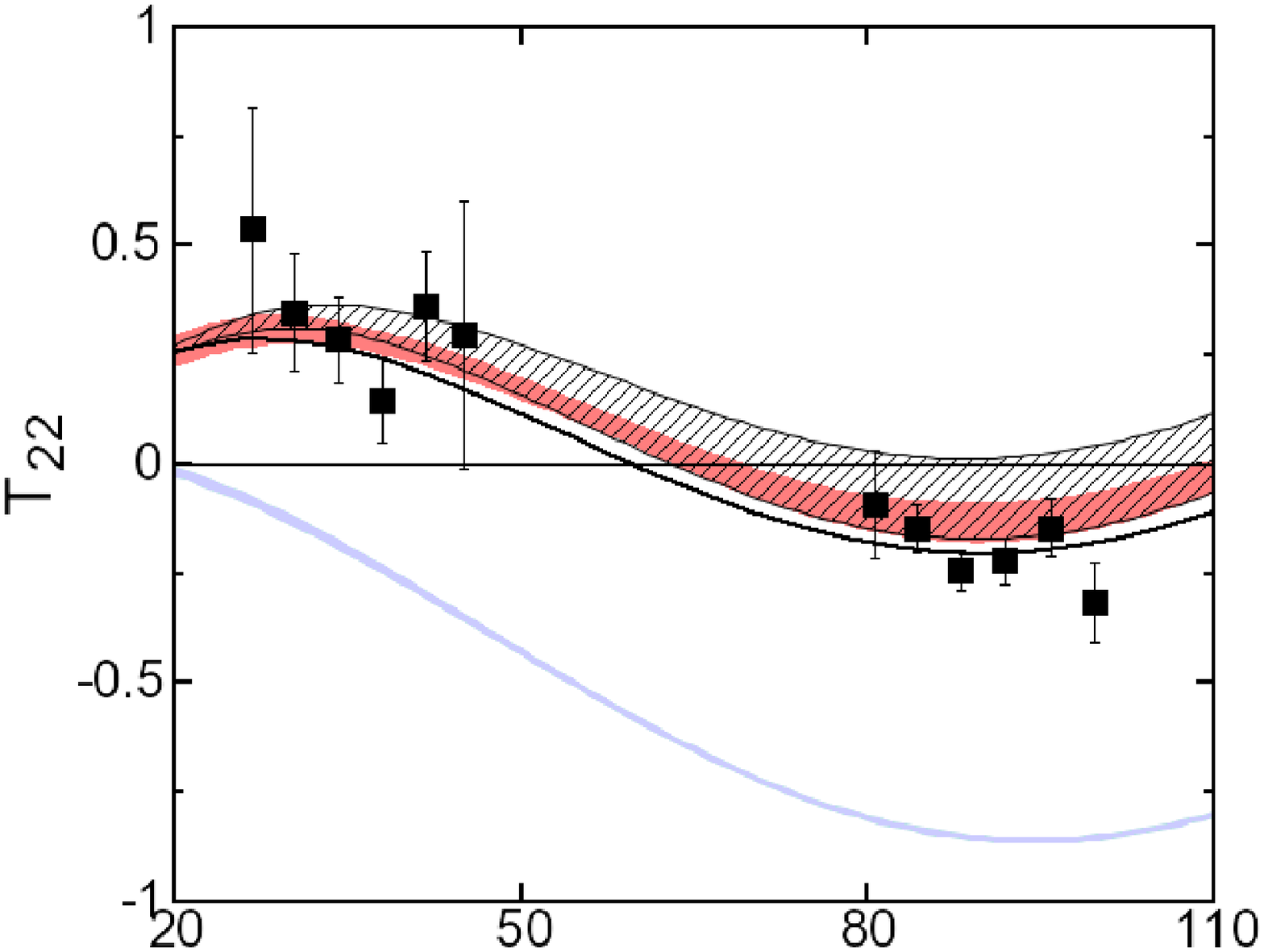} \\
     \hspace{-3mm}\includegraphics[width=6.0cm,height=5.2cm]{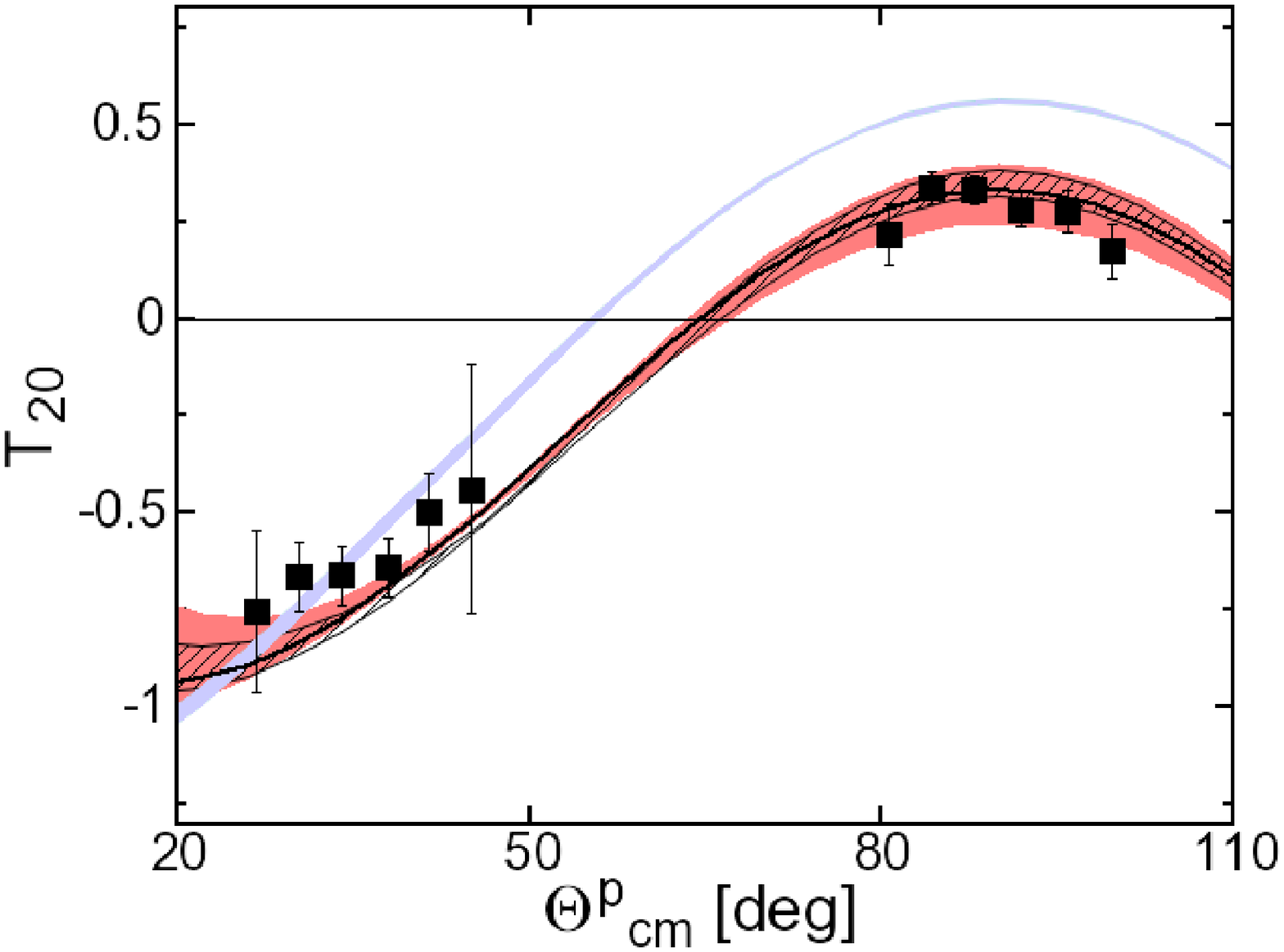} &
     \hspace{-1mm}\includegraphics[width=6.0cm,height=5.2cm]{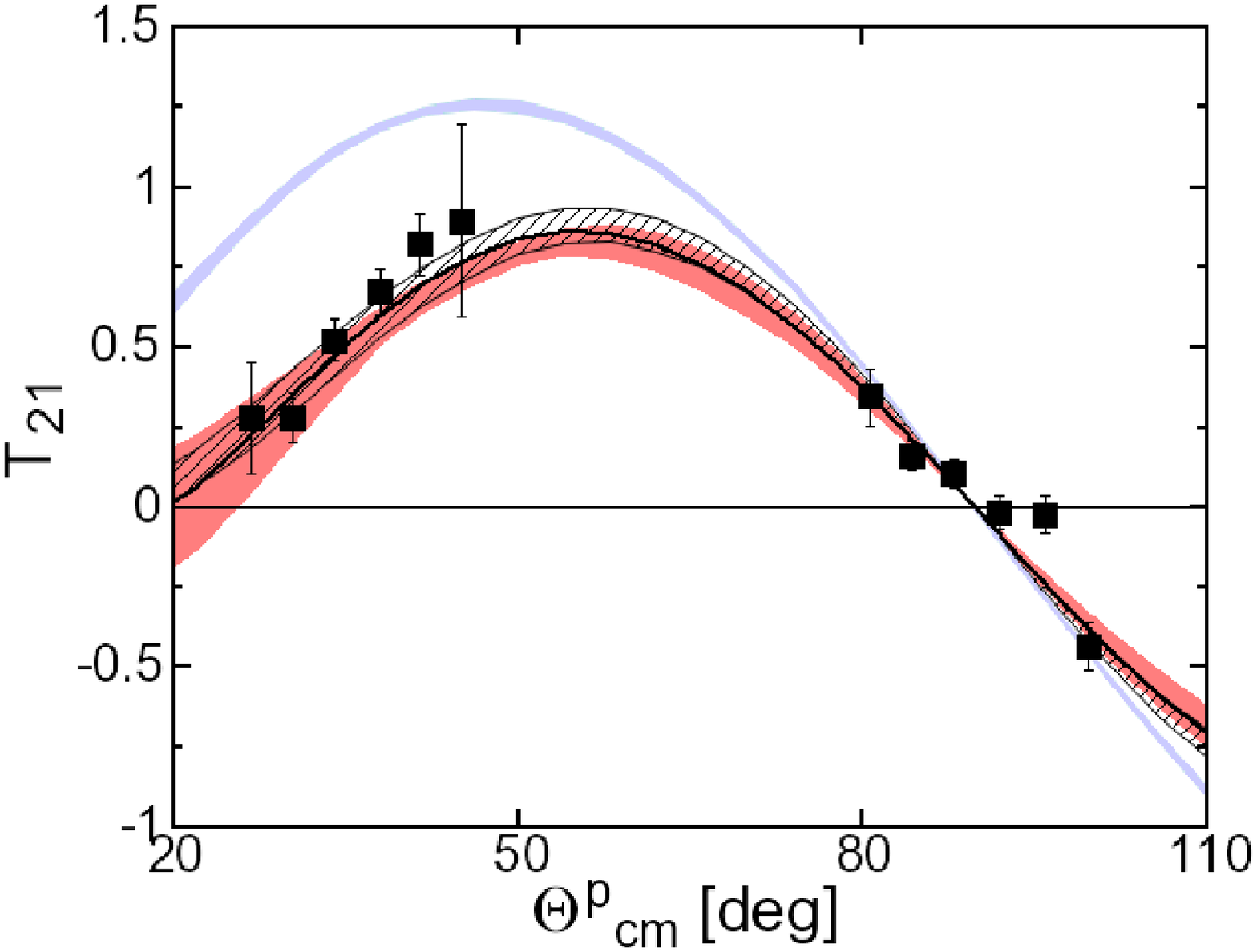} &
     \hspace{-1mm}\includegraphics[width=6.0cm,height=5.2cm]{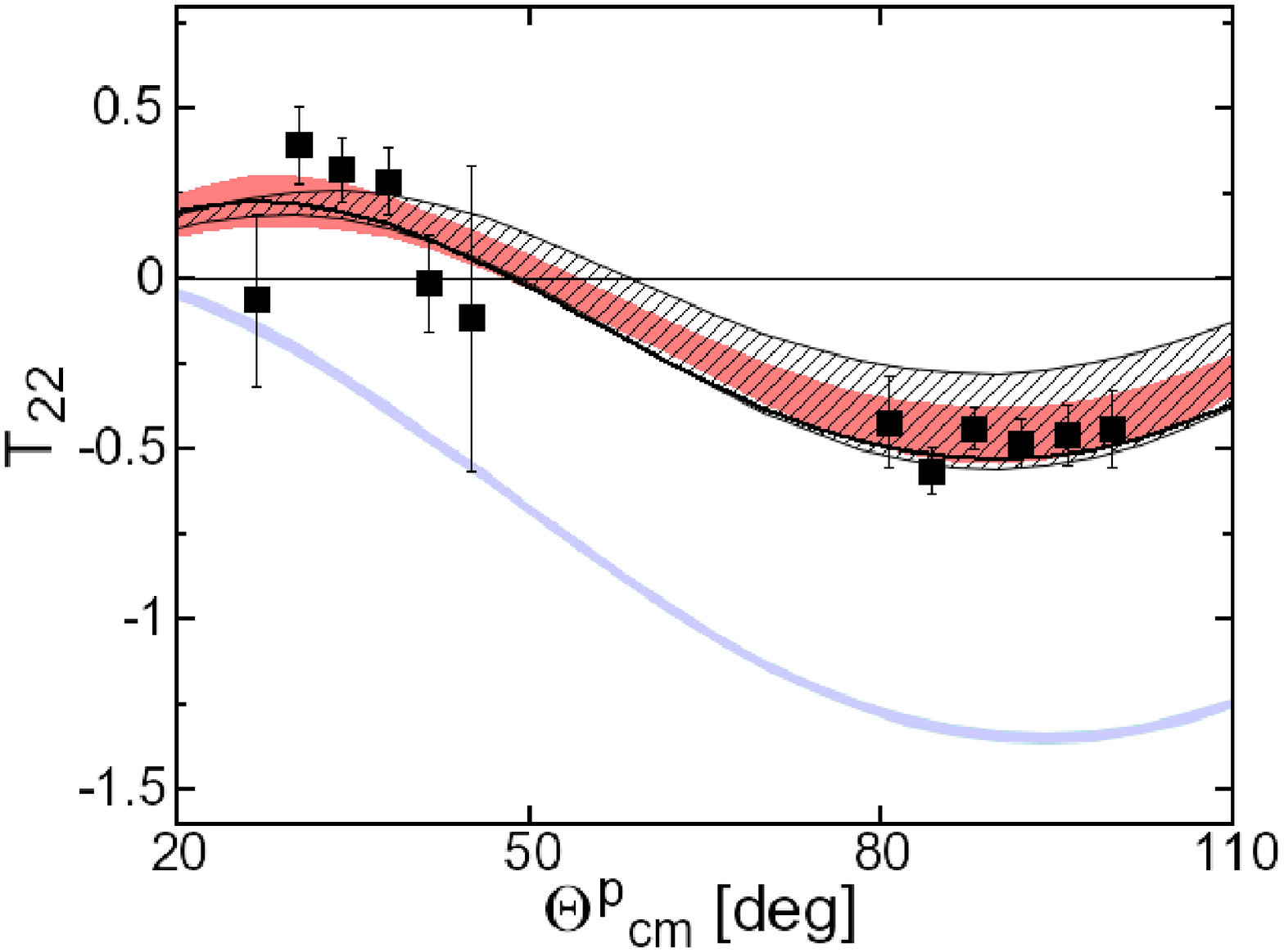} \\ 
    \end{tabular}
\caption{\label{fig2}(color online)
Deuteron tensor analyzing powers vs. proton emission angle for two $E_\gamma$- energy bins.
The upper row shows results for bin energy $E_\gamma$ = 25-45 MeV. The lower row shows results for bin energy $E_\gamma$ = 45-70 MeV.
The bands and lines have the same meaning as in Fig.~\ref{fig1}.
The experimental data are from Rachek et al.~\cite{rachek:2007}.}
\end{figure}

We have also calculated the deuteron tensor analyzing powers as a function of the proton emission
angle for two $E_{\gamma}$ - energy bins. Here, we focus on a comparison of
our calculations with recent experimental 
tensor analyzing powers $T_{2q}$ for low energies from~\cite{rachek:2007} and
do not show the results for vector analyzing power $iT_{11}$ 
as there exist no experimental data for this observable. In order to be able
to compare the theoretical calculations with the data from~\cite{rachek:2007}, 
our predictions for the exclusive observables have been integrated over the
relevant intervals of the initial photon energy and 
angular regions. In Fig.~\ref{fig2} the results for the angular distribution
at the bin energies of $E_{\gamma}=25-45$ MeV and $E_{\gamma}=45-70$ MeV
together with experimental data are presented.
For all deuteron tensor analyzing powers one observes a rather good agreement
between the AV18 potential prediction, chiral results and
experimental data. 
The effects of the TPE contributions turn out to be very small. 
Also, no broadening of the bands with increasing photon energy is observed.
All this suggests that the deuteron tensor analyzing powers are driven by 
the long-range parts of the current and are not sensitive to the short-range
contributions. We further emphasize some disagreement with the data for
$T_{21}$ at forward angles. In the future, it would be interesting to see
whether these conclusions are affected by the inclusion of the subleading 
OPE terms in the current operator. 

\begin{figure}[t]
\begin{center}
    \begin{tabular}{lr}
     \hspace{0mm}\includegraphics[width=7.0cm,height=5.7cm]{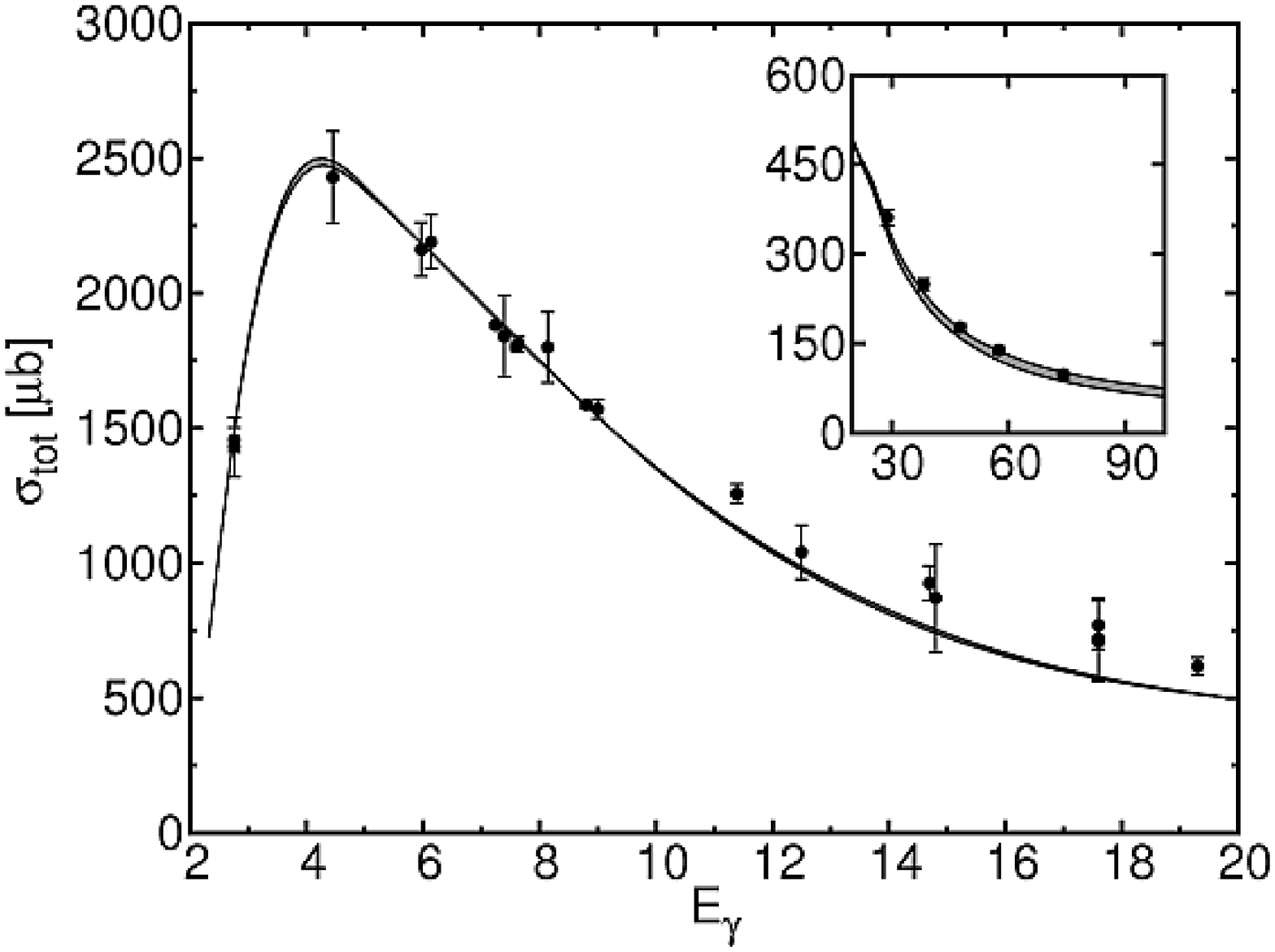}
     \hspace{0mm}\includegraphics[width=6.8cm,height=5.7cm]{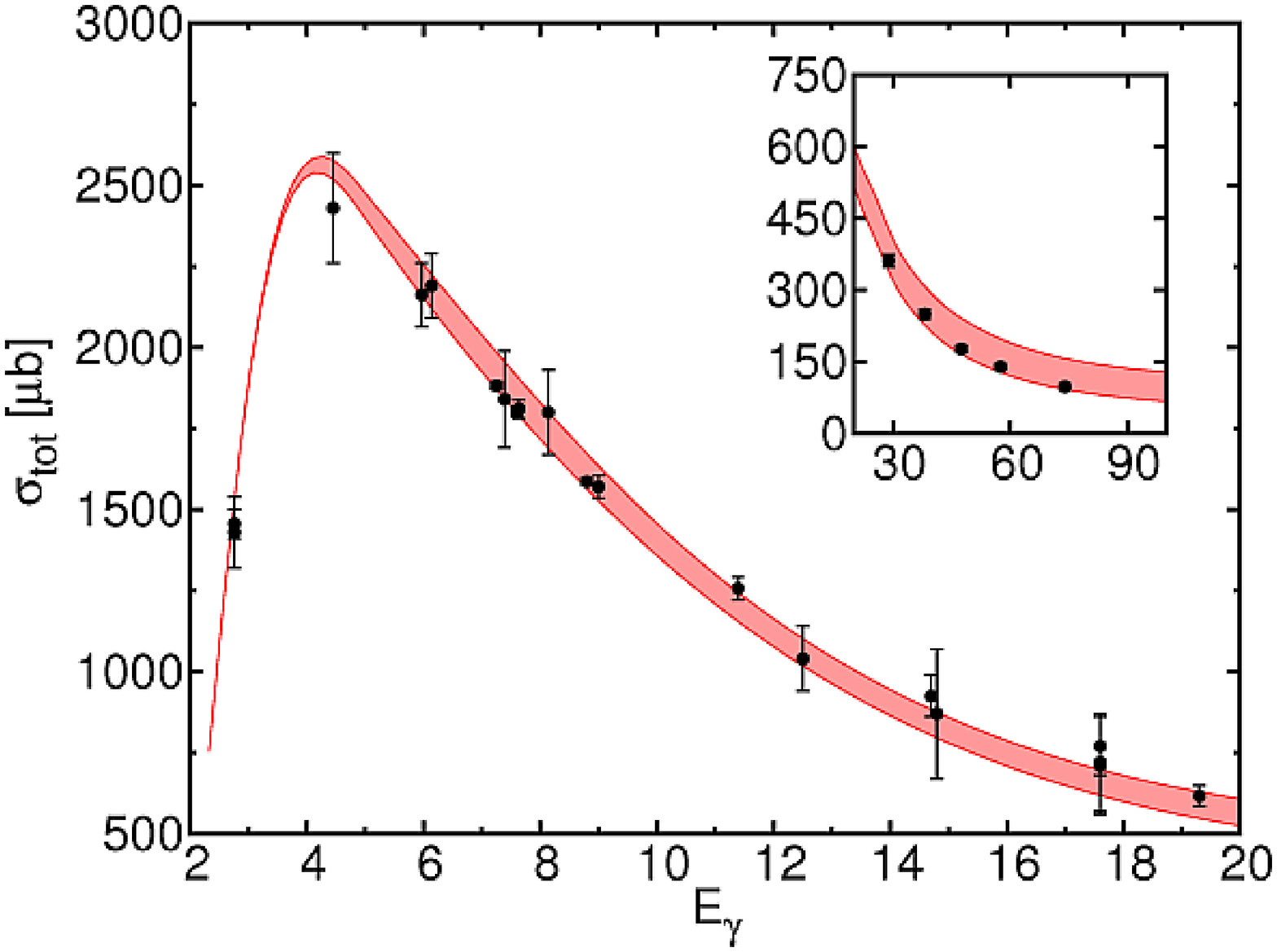}
     \end{tabular}
\end{center}
\vspace{-3mm}
\caption{\label{totgd}(color online) Total cross section for photodisintegration of 
the deuteron as a function of photon energy beam.
In the left panel results for the single nucleon current and OPE contribution are shown. 
In the right panel, the results obtained with an additional TPE currents are given.
The experimental data are the same as in Ref.~\cite{arenhovel:1991}.} 
\end{figure}

Finally, we have also calculated the total cross section. This observable was
extensively studied in many publications using various theoretical 
approaches, see e.g.~\cite{arenhovel:1991}.
In Fig.~\ref{totgd}, the total cross section for $E_\gamma\approx$ 2--80 MeV is presented.
The experimental data are taken from
Refs.~\cite{arenhovel:1991,gilman:2002}. In the left panel, we see the results
for the single nucleon and OPE current.  
For this particular case the width of the prediction band is negligible. The
theoretical predictions agree rather well with the experimental data.
The right panel in this figure shows that the 
effects of the TPE currents are clearly visible, especialy at higher 
photon energies. We also notice that the band width increases significantly 
once the TPE contributions are included. 

To conclude, we observe that for all considered observables the 2N current
operator plays an important role
and a restriction to the single nucleon current operator leads to a
strong disagreement with the data. The inclusion of the leading OPE current
is absolutely necessary to achieve a decent description  of the experimental
data. The effects of the TPE current are clearly visible in the differential
cross section and some polarization observables such as $\Sigma^1$ and
$P_y$. One also observes a rather good agreement
between the results based on chiral EFT and the AV18 potential combined with
the corresponding current operators. 

\section{Two-pion exchange currents in the 3N system}
\def\theequation{\arabic{section}.\arabic{equation}}
\setcounter{equation}{0}
For 3N reactions, we use the framework and its numerical implementation
described in detail in~\cite{golak:2005}.
In this work, we will only briefly introduce the key points focusing mainly on the
current operator in the 3N system.
The starting point is exactly the same as for the 2N reaction.
We consider the general matrix element of the current operator between the 3N
bound state,  $|\Psi^{3N}_{\rm bound}\rangle$,
and scattering state $|\Psi^{3N}_{\rm scatt}\rangle$ for the 3N system
\begin{equation}
\label{N3N}
 N^{\mu}\equiv \langle \Psi^{3N}_{\rm scatt}|J^{\mu}(\vec Q)|\Psi^{3N}_{\rm bound}\rangle \,.
\end{equation}
The 3N bound state $|\Psi^{3N}_{\rm bound} \rangle$ is obtained in the standard way 
from the appropriate Faddeev equation~\cite{naszbs2}.
The current operator $J^{\mu}(\vec Q)$ acts effectively between the internal initial and final 3N states.
These internal states are conventionally expressed in the momentum space
in terms of two Jacobi momenta, $\vec p$ and $\vec q$~\cite{gloekle:1983}.
The momentum $\vec p$ describes
a relative motion within a 2N subsystem
(here we choose the subsystem consisting of nucleons 2 and 3). The
momentum $\vec q$ describes the motion of the spectator nucleon (here nucleon 1)
with respect to that 2N subsystem
\begin{equation}
\vec p = \frac12 \Bigl( \vec p_2 - \vec p_3 \Bigr) \, , \quad \quad 
\vec q = \frac23 \Bigl( \vec p_1 - \frac12 \bigl( \vec p_2 + \vec p_3 \bigr) \Bigr) \, .
\label{jacobiq}
\end{equation}
We consider two types of 3N scattering states. In the first case
two nucleons bound in the deuteron emerge with the accompanying third nucleon
and the asymptotic motion
of this unbound nucleon is described by the Jacobi momentum $\vec q_0$.
In the second case, we have three free nucleons in the final state
and their asymptotic relative motions are represented by $\vec p$ and $\vec q$.

In order to calculate the crucial matrix elements $N^{\mu}$ given in~(\ref{N3N}),
it is not necessary to solve the corresponding Faddeev equations directly
for the 3N scattering states~\cite{golak:2005}. Instead, we solve a Faddeev-type equation for an auxiliary state $\mid U \rangle $. In our calculations, we do not include the effects of 3N forces.
Thus, the equation for $\mid U \rangle $ takes a simpler form
\begin{equation}
\label{U3n}
\mid U \rangle =
t G_0 \, ( 1 + P ) J^{\mu}(\vec Q)\mid \Psi^{3N}_{\rm bound} \rangle
+ t G_0 P \mid U \rangle  \,.
\end{equation}
The corresponding nuclear matrix elements are then given by 
\begin{eqnarray}
\label{npn}
N^{\mu}_{Nd} &=&
\langle \Phi_d \mid ( 1 + P ) J^{\mu}(\vec Q)\mid \Psi^{3N}_{\rm bound} \rangle
+ \langle \Phi_d \mid P \mid U \rangle \,, \nonumber  \\
N^{\mu}_{Npn} &=&
\langle \Phi_0 \mid ( 1 + P ) J^{\mu}(\vec Q)\mid \Psi^{3N}_{\rm bound} \rangle
+ \langle \Phi_0 \mid ( 1 + P ) \mid U \rangle  \, ,
\end{eqnarray}
with NN $t$-matrix in the 3N space, $G_0$ the free 3N
propagator and $P$  the permutation
operator $P= P_{12} P_{23} + P_{13} P_{23}$. Further, $|\Phi_d\rangle$ is the 
antisymmetrized product state containing the 
deuteron and the momentum state for the relative motion of the third nucleon. Finally, $|\Phi_0\rangle$ is the antisymmetrized
product state describing the two relative motions among the three outgoing nucleons. 
For details about the solution of Eqs~(\ref{U3n})--(\ref{npn}), see Ref.~\cite{golak:2005}. 
It is important to mention that these
equations are solved in the partial wave basis. In the following, we, therefore, briefly discuss
the partial wave decomposition of the current operator which can generally be
written in the form 
\begin{equation}
J^\mu = J^\mu(1)+ J^\mu(2) + J^\mu(3)+ J^\mu(2,3)+J^\mu(3,1)+J^\mu(1,2).
\end{equation}
There are three pairs in the 3N system, but it is sufficient to include a
contribution just from one pair exploiting the fully antisymmetric nature of
the 3N states. The two-nucleon current operator $J^\mu(2,3)$ is defined according to Eq.~(2.5).
Compared to Eq.~(\ref{J2PI2N}), in the case of the 3N system, we have
additional contributions from the $T_1$ isospin structure.  
Thus, the non-vanishing contributions emerge from
\begin{equation}
\vec{J}_{2\pi}(2, 3) = \sum_{\beta=3}^{10} f_{1}^{\beta}(\vec{q}_2,\vec{q}_3) T_{1} \vec{O}_{\beta} 
+ \sum_{\beta=3}^{10}f_{2}^{\beta}(\vec{q}_2,\vec{q}_3) T_{2} \vec{O}_{\beta} 
+ f_{3}^{2}(\vec{q}_2,\vec{q}_3) T_3\vec{O}_{2}\,,
\end{equation}
where $\vec{q}_2$ and $\vec{q}_3$ are the momentum transfers of nucleons 2 and 3, respectively. 
We utilize the so-called $jI$ coupling scheme for the 3N basis states
\begin{equation}
\mid p \, q \, \alpha \rangle =
\Bigl| p \, q \, \bigl(l s \bigr) j \Bigl(\lambda \frac12 \Bigr) I  \bigl( j I \bigr) J M
\Bigr\rangle \Bigl| \Bigl( t \frac12 \Bigr) T M_T \Bigr\rangle
\equiv \mid p \, q \, \alpha_J \rangle \mid \alpha_T \rangle \,.
\label{per103}
\end{equation}
Here, $l$, $s$, $j$ and $t$ refer to the orbital angular momentum, spin, total
angular momentum and isospin of the (2-3) subsystem, respectively. 
The angular momentum of nucleon 1 is coupled with its spin $\frac12$ to the total
angular momentum $I$. Finally, the subsystem total angular momentum $j$ is coupled with $I$
to give the total 3N angular momentum $J$ with the projection $M$. Similar coupling
in the isospin space leads to the total 3N isospin $T$ with the corresponding magnetic
quantum number $M_T$.

Analogously to the procedure described in section II, we compute 
the general matrix element of the 2N current in the 3N basis
\begin{eqnarray}
\left\langle  p' q' \alpha ' \left|\vec J_{2\pi}( 2, 3 )
\right| p q \alpha \right\rangle &=& \langle p' q' \alpha_{J'}|\vec O_\beta \,f^{\beta}_{\eta}(\vec{q_{2}},\vec{q_{3}})|p q \alpha_J\rangle \,  
\langle \alpha_{T'}|T_{\eta}|\alpha_T\rangle
\nonumber \\
&=&
\sum_{m_j , \,  m_{j'}} C(j' I' J' ; m_{j'} , M' -m_{j'}, M') \,
C(j I J;m_j , M-m_j, M)
\nonumber \\
&\times &
I_{23} ( p',p,Q; (l's')j' m_{j'} ,(ls)j m_j )
\nonumber \\
&\times&I_1\left( q',q,Q ; \left(\lambda' \frac{1}{2}\right) I' M'-m_{j'} ,\left( \lambda \frac{1}{2}\right) I M- m_j \right)
\left\langle \alpha _{T'} \left| T_\eta \right| \alpha_T \right\rangle
\label{per107}
\end{eqnarray}
with
\begin{eqnarray}
I_{23} (  p',p,Q; (l's')j' m_{j'} ,(ls)j m_j)
&=& \int d\hat  {p'}\int d\hat {p} \sum\limits_{m_{l'}}\sum\limits_{m_l} C (l' s' j' ;m_{l'}, m_{j'} - m_{l'}, m_{j'} )
Y_{l',m_{l'}}^* (\hat {p'})  \\
&\times& C (l s j ; m_l , m_j - m_l , m_j ) Y_{l,m_l} ( \hat {p})\,
f^{\beta}_{\eta}(\vec{q_{2}},\vec{q_{3}})\, \langle s'\,m_{j'}-m_{l'}|\vec
O_\beta|s\,m_j-m_l \rangle \nonumber 
\end{eqnarray}
and
\begin{equation}
I_1\left( q',q,Q ; \left(\lambda' \frac{1}{2}\right) I' M'-m_{j'} ,\left( \lambda \frac{1}{2}\right) I M- m_j \right)
=
\int d\hat {q'} {\cal Y }^{I', M'-m_{j'}\,*}_{\lambda'\frac{1}{2}} \left(\hat{q'}\right)
\frac{\delta ( q -\vert \vec {q'} + \frac{1}{3} \vec Q \vert ) } {q ^2} 
{\cal Y }^{I, M-m_j}_{\lambda\frac{1}{2}}  \left( \widehat {\vec {q'} + \frac{1}{3} \vec Q} \,\right)\,,
\end{equation}
where we have introduced 
\begin{eqnarray}
{\cal Y}^{I\,\nu}_{\lambda\frac{1}{2}}(\hat q ) \equiv
\sum\limits_{m} C \left(\lambda \frac{1}{2} I ; m, \nu-m, \nu \right)
Y_{\lambda,m} (\hat q) \left| \frac{1}{2} \, \nu-m \right\rangle .
\end{eqnarray}
For $I_{23}$, we recognize the same type of a matrix element
we dealt with in the 2N space. Now, however, the isospin part is separated
out. This is because there are much more isospin combinations in the 3N system
compared to processes on the deuteron which has the total isospin zero. 
This separation allows us, in particular, to calculate $I_{23}$ once and use it
both for the reaction on $^3$He and $^3$H. For the numerical implementation it
is, however, still important 
to use the properties of the matrix elements $\left\langle \alpha _{T'} \left|
    T_{\eta} \right| \alpha_T \right\rangle$ (${\eta=1, 2, 3}$) in order to
reduce the number of necessary four-fold integrals in $I_{23}$, even if the isospin dependence
is now treated separately. Below we give the matrix elements of the three isospin operators
in the 3N isospin space. As already mentioned, we assume that the operators act on
the 3N bound state ($^3$He or $^3$H), which has the total isospin $T=1/2$.
It is then straightforward to obtain 
\begin{eqnarray}
\left\langle \, \left( t' \frac12 \right) T' m_{T'}  \Bigl|
{\: T_1 }
\Bigr| \left( t \frac12 \right) \frac12 m_T \, \right\rangle &\equiv& 
\left\langle \, \left( t' \frac12 \right) T' m_{T'}  \Bigl|
{\: {\left( {\vec{\tau}(2)} + {\vec{\tau}(3)} \right) }_3}
\Bigr| \left( t \frac12 \right) \frac12 m_T \, \right\rangle \nonumber \\
&=&
C\left( 1,\frac12,T';0, m_T, m_{T'} \, \right) \, 
\sqrt{12} \, \sqrt{ (2 t' + 1) (2 t + 1)} \,
\left\{ \begin{array}{ccc}
1 & t & t' \\ 
\frac12 & T' & \frac12
\end{array}\right\}
\,
\left\{ \begin{array}{ccc}
1 & \frac12 & \frac12 \\ 
\frac12  & t' & t
\end{array}\right\} \nonumber \\
&\times & (-1)^{t+t'+\frac12+T'} \, \left( 1 + (-1)^{t+t'} \, \right) \, ,\\[15pt]
\left\langle \, \left( t' \frac12 \right) T' m_{T'}  \Bigl|
{\: T_2 }
\Bigr| \left( t \frac12 \right) \frac12 m_T \, \right\rangle & \equiv &
\left\langle \, \left( t' \frac12 \right) T' m_{T'}  \Bigl|
{\: {\left( {\vec{\tau}(2)} - {\vec{\tau}(3)} \right) }_3}
\Bigr| \left( t \frac12 \right) \frac12 m_T \, \right\rangle\nonumber \\
&=&
C\left( 1,\frac12,T';0, m_T, m_{T'} \, \right) \,
\sqrt{12} \, \sqrt{ (2 t' + 1) (2 t + 1)} \,
\left\{ \begin{array}{ccc}
1 & t & t' \\ 
\frac12 & T' & \frac12
\end{array}\right\}
\,
\left\{ \begin{array}{ccc}
1 & \frac12 & \frac12 \\ 
\frac12  & t' & t
\end{array}\right\} \nonumber \\
&\times & (-1)^{t+t'+\frac12+T'} \, \left( 1 - (-1)^{t+t'} \, \right) \, ,
\end{eqnarray}
\begin{eqnarray}
\left\langle \, \left( t' \frac12 \right) T' m_{T'}  \Bigl|
{\: i T_3 }
\Bigr| \left( t \frac12 \right) \frac12 m_T \, \right\rangle &\equiv& 
\left\langle \, \left( t' \frac12 \right) T' m_{T'}  \Bigl|
{i \: {\left( {\vec{\tau}(2)} \times {\vec{\tau}(3)} \right) }_3}
\Bigr| \left( t \frac12 \right) \frac12 m_T \, \right\rangle \nonumber \\
&=&
C\left( 1,\frac12,T';0, m_T, m_{T'} \, \right) \, 
12 \sqrt{3} \, \sqrt{ (2 t' + 1) (2 t + 1)} \,
\left\{ \begin{array}{ccc}
1 & t & t' \\ 
\frac12 & T' & \frac12
\end{array}\right\}
\,
\left\{ \begin{array}{ccc}
1 & 1 & 1 \\ 
\frac12 & \frac12 & t \\ [2pt]
\frac12  & \frac12 & t'
\end{array}\right\}\nonumber \\
&\times & (-1)^{1+t+\frac12+T'} \, .
\end{eqnarray}

\section{Results for photodisintegration of $^3$He}
\def\theequation{\arabic{section}.\arabic{equation}}
\setcounter{equation}{0}
We are now in the position to discuss our results for two- and three-body
photodisintegration of $^3$He at three example photon laboratory energies $E_\gamma$ =
12, 20.5 and 50 MeV. 
The Coulomb force between two-protons in three-nucleon scattering states is not taken into account.
The three-nucleon matrix elements  $\vec N$ are obtained using the partial wave decomposition,
with the total angular momentum of the three-nucleon system $J\le 15/2$ and including all 
partial waves with the subsystem total angular momentum $j\le 3$. 
The nuclear matrix elements $\vec N$ are computed as using  the formalism described in section IV.
Given $\vec N$, one can calculate cross sections and polarization
observables which are expressed in terms of  the nuclear matrix elements with different
spin projections carried by the initial photon, the $^3$He nucleus and 
the outgoing nucleons and/or deuteron.
For more details we refer the reader to Refs.~\cite{romek1,golak:2005}.  
In the following we just present our sample results for the chiral EFT approach.

\begin{figure}[t]
  \begin{center}
    \begin{tabular}{lr}
     \includegraphics[width=5.8cm,height=5.0cm]{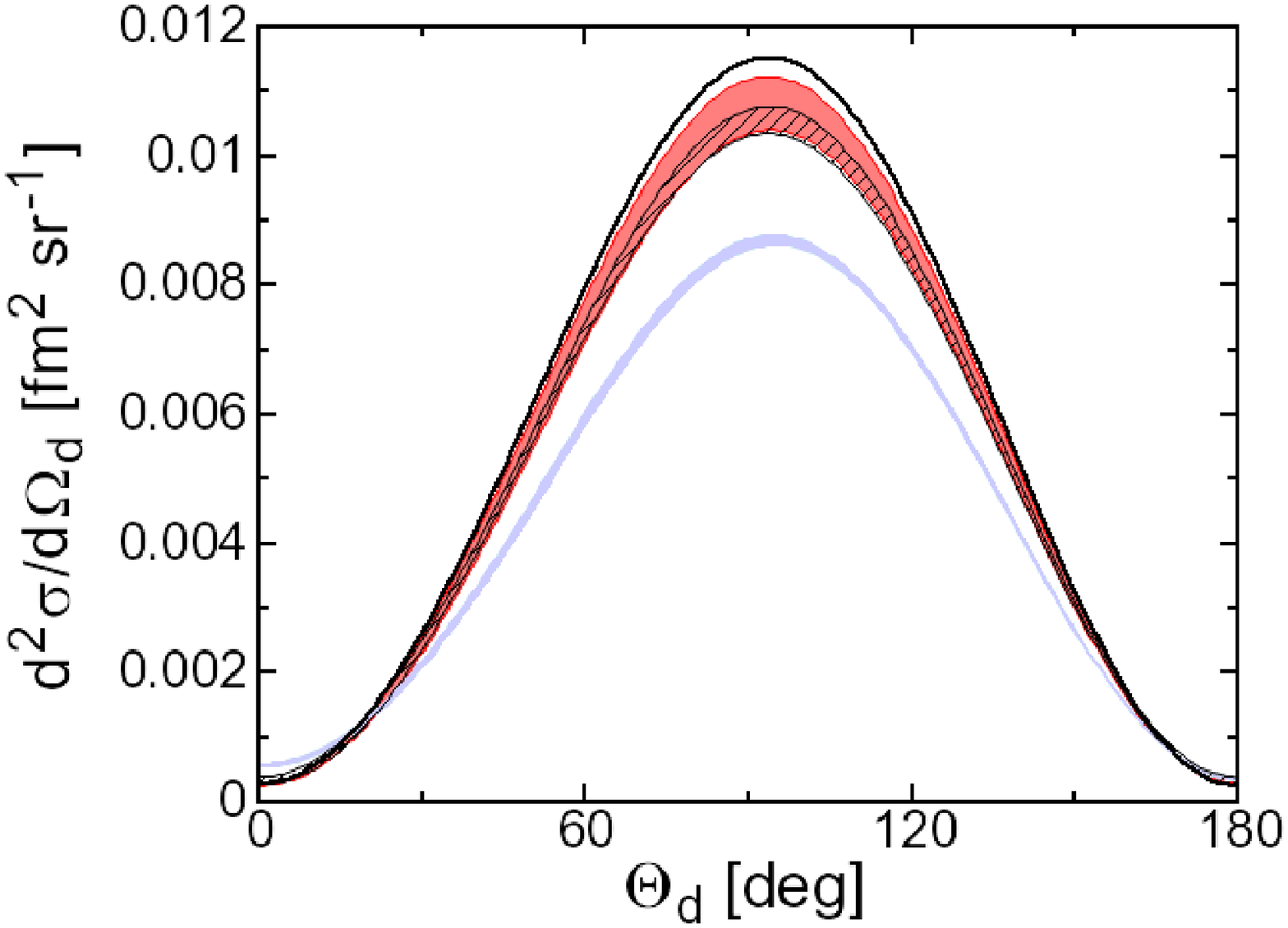}
     \includegraphics[width=5.8cm,height=5.0cm]{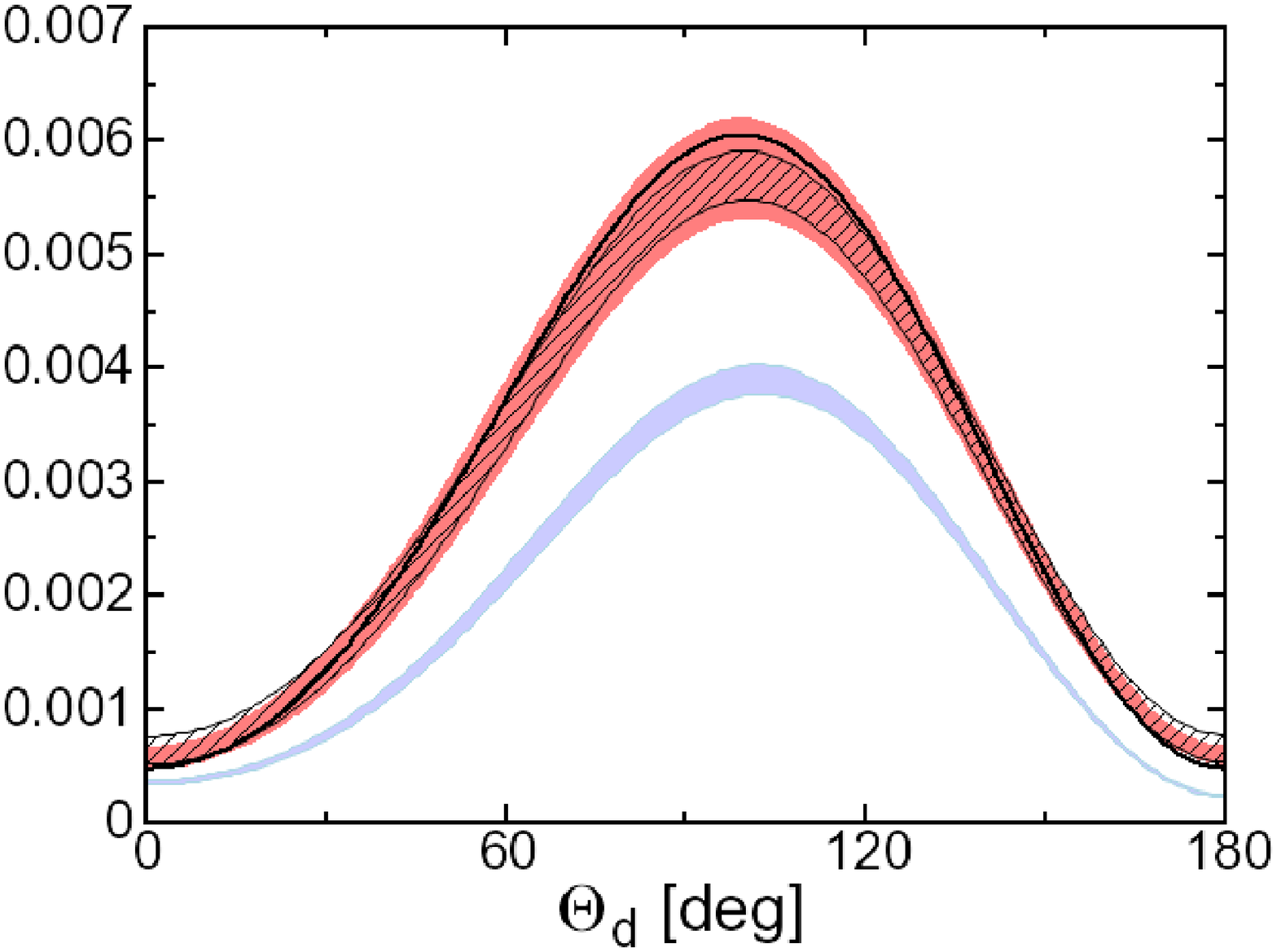}
     \includegraphics[width=5.8cm,height=5.0cm]{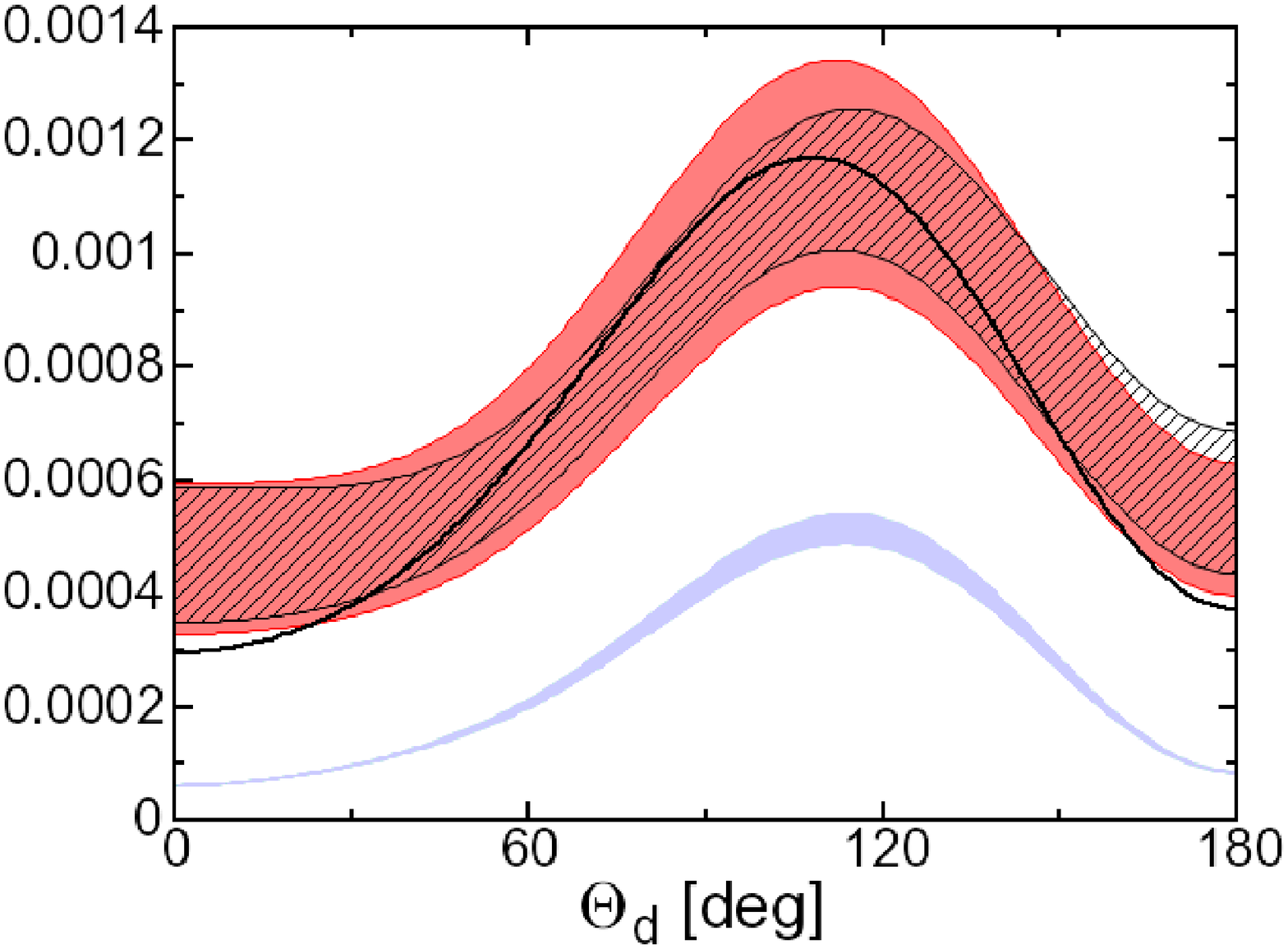}
    \end{tabular}
  \end{center}
  \caption{\label{fg3n1}(color online) Differential cross section in the
    laboratory frame for $^3$He two-body photodisintegration 
at the photon laboratory energies $E_{\gamma}=12$ MeV (left), $E_{\gamma}=20.5$
MeV (middle)  and $E_{\gamma}=50$ MeV (right). 
The band covers N$^2$LO chiral predictions for different cut-off parameter
values. The light (blue) band covers results obtained with the single-nucleon
current. 
In the case of the hatched band, the current operator is taken as a sum of the
single nucleons current  and
one-pion exchange current. 
The dark (pink) band covers N$^2$LO chiral predictions for different cut-off
parameter values and the current operator is taken as a sum of the single nucleons
current, one-pion exchange current and two-pion exchange current.
The solid line represents predictions obtained with the AV18 nucleon-nucleon
potential and the related exchange currents~\cite{golak:2005}.}
\end{figure}

We begin with the exclusive unpolarized cross section for two-body breakup of $^3$He,
$d^2\sigma/d \Omega_d$, where the final deuteron would be observed.
It is depicted in Fig.~\ref{fg3n1} as a function of
the deuteron scattering angle $\theta_{d}$ defined with respect to the initial photon direction
at the photon laboratory energies $E_{\gamma}$ = $12$, $20.5$ and $50$ MeV.
We observe a similar behaviour as compared to the differential cross section
in photodisintegration of the deuteron. 
The single nucleon current contribution yields significantly lower values as
compared to the ones which include MECs.  
The TPE bands overlap with the OPE bands and appear to be broader than the OPE bands.
As expected, the bands become wider with increasing photon energy.  

\begin{figure}[t]
\begin{center}
 \begin{tabular}{lr}
   \hspace{-2.5mm}\includegraphics[width=6.2cm,height=4.8cm]{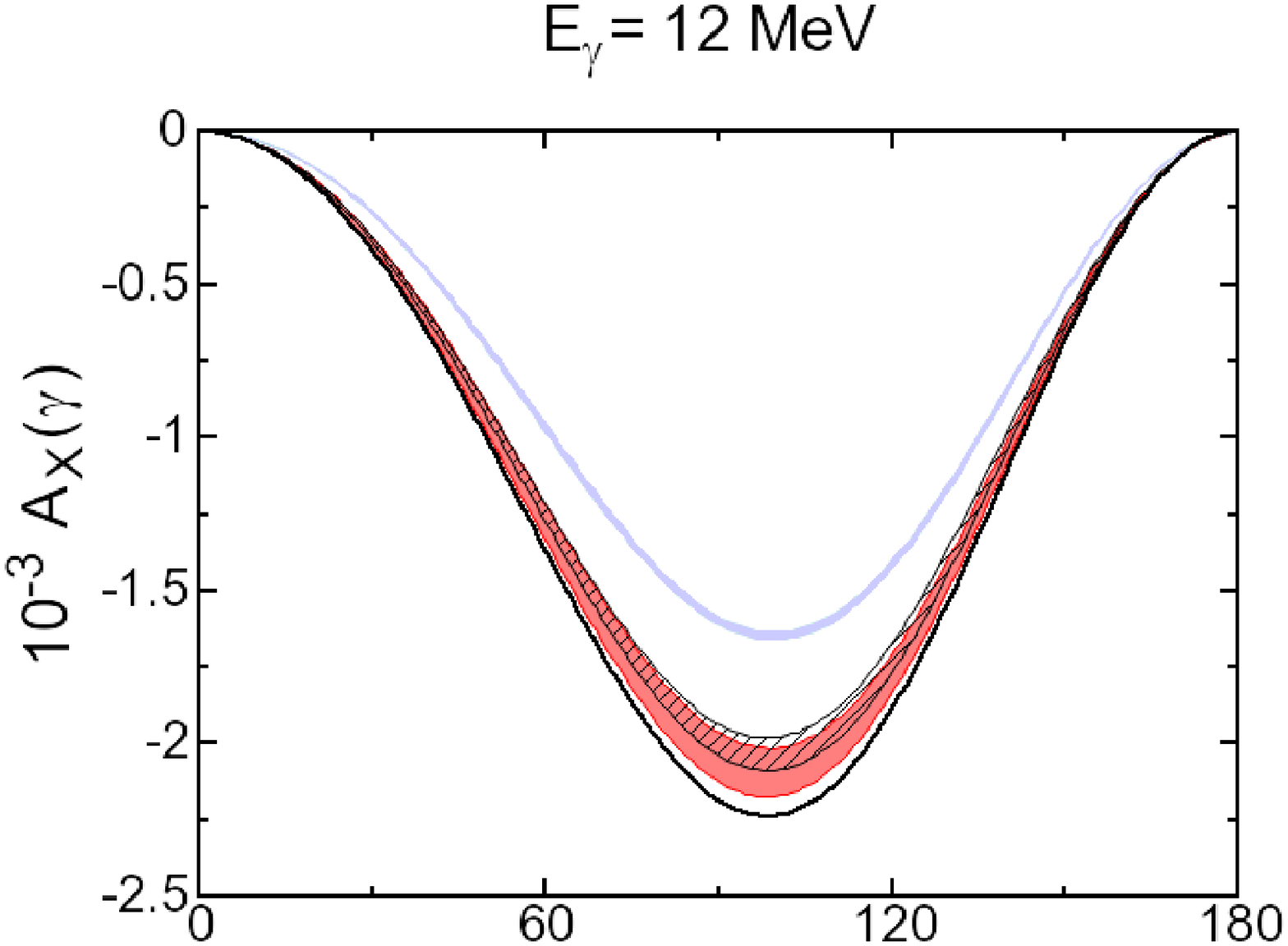}
   \hspace{-1mm}\includegraphics[width=5.9cm,height=4.8cm]{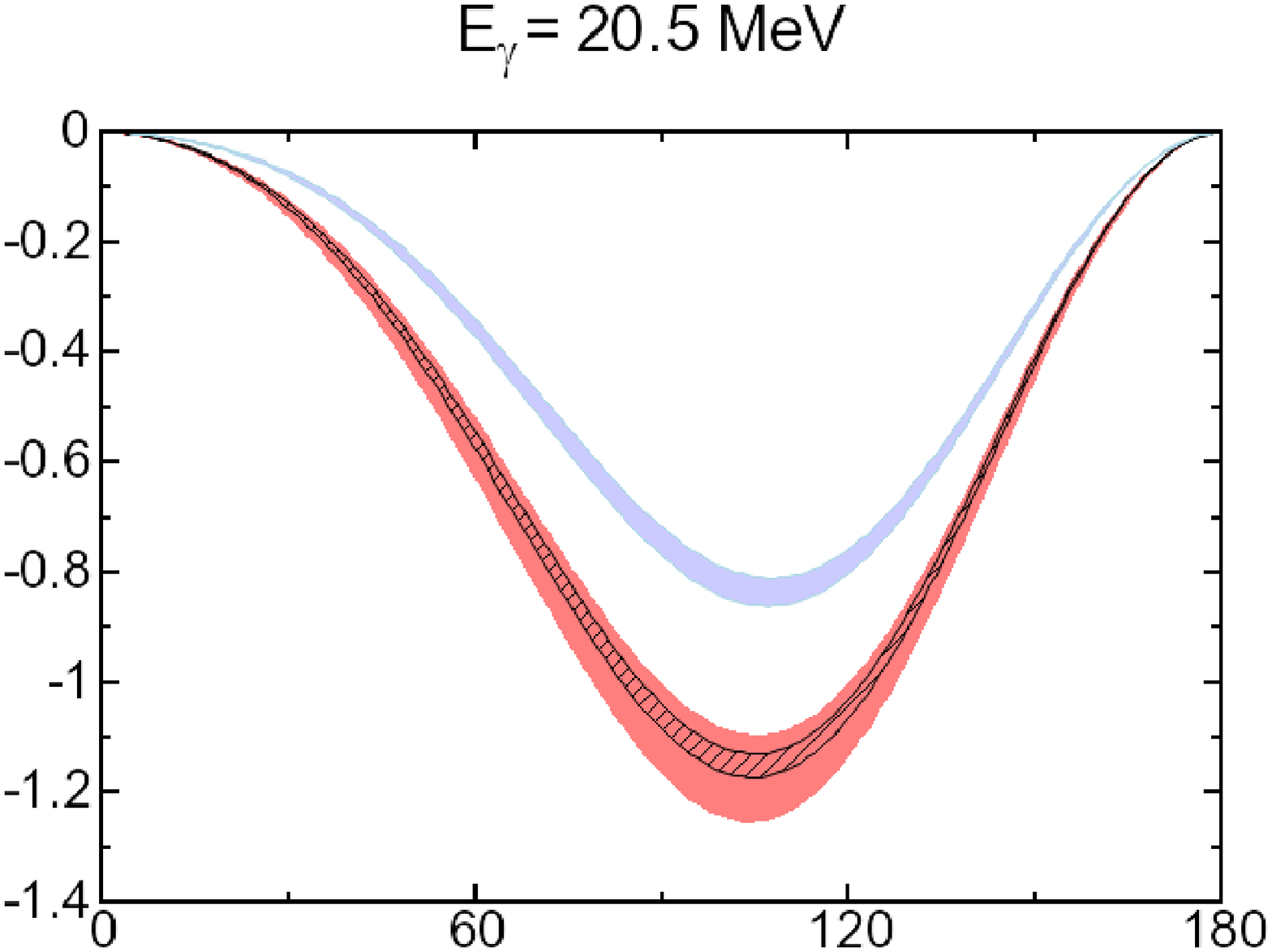}
   \hspace{-1.5mm}\includegraphics[width=5.8cm,height=4.8cm]{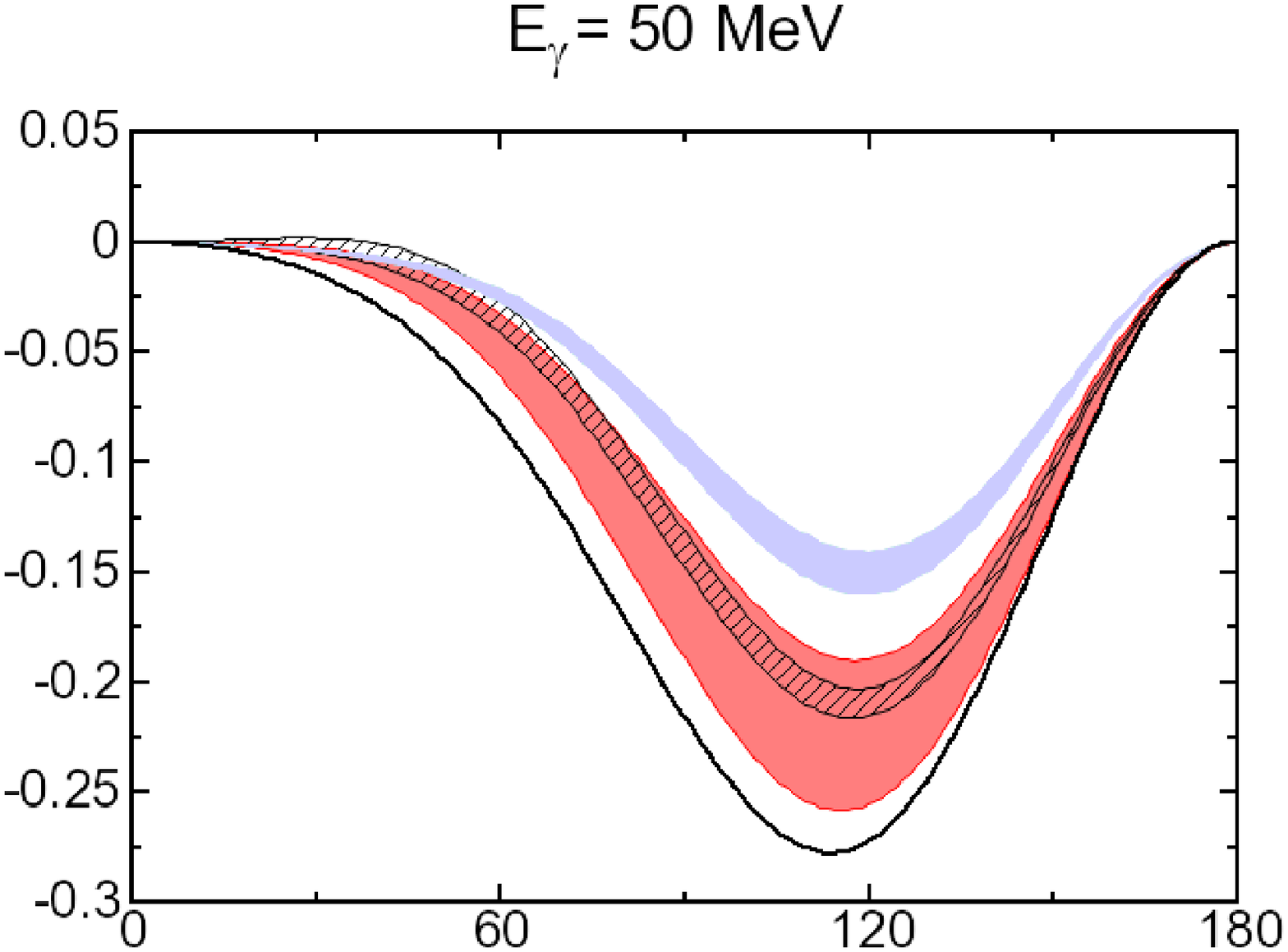}\\
   \hspace{-1mm}\includegraphics[width=6.1cm,height=4.3cm]{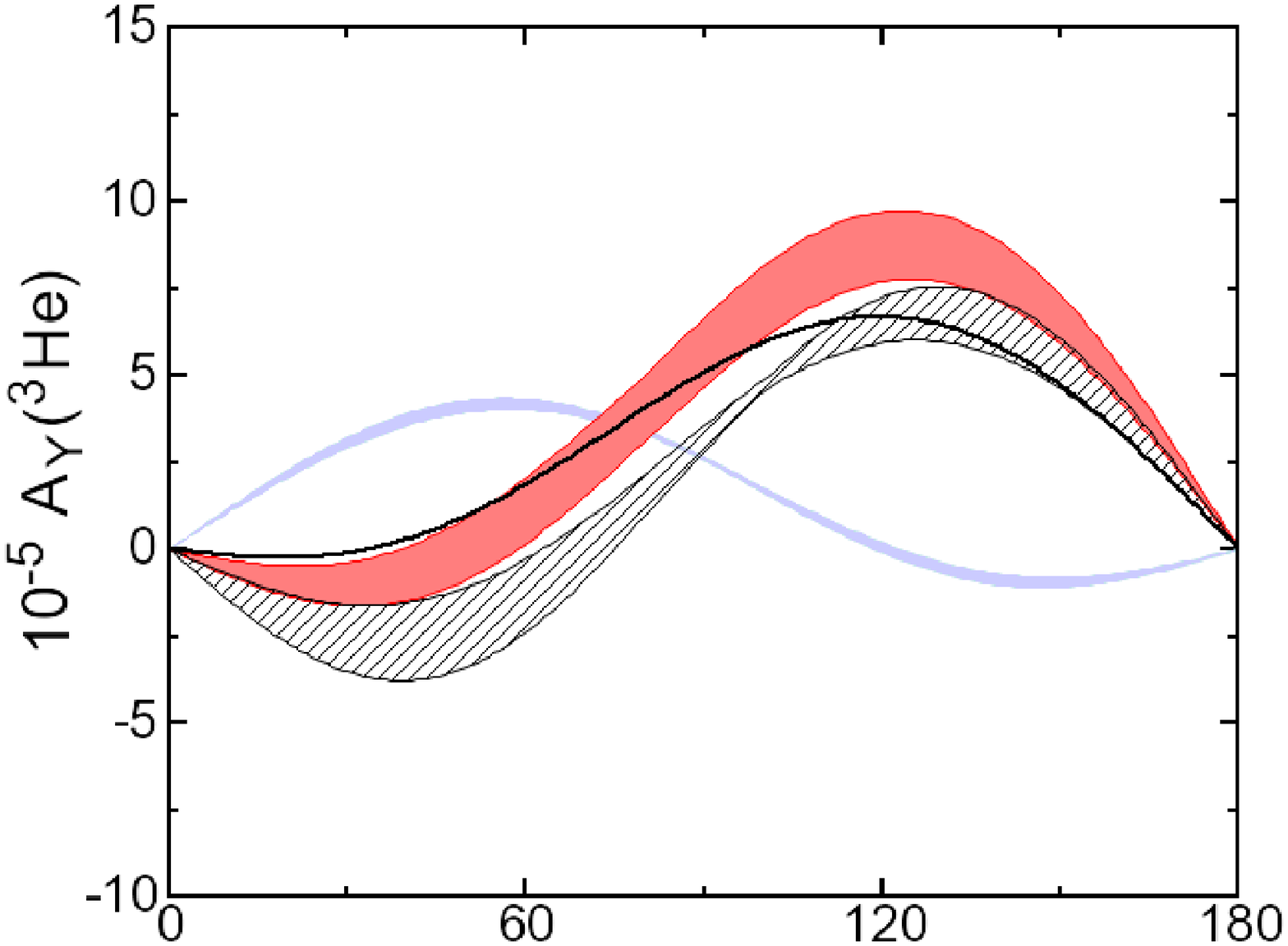}
   \hspace{-0.5mm}\includegraphics[width=5.9cm,height=4.3cm]{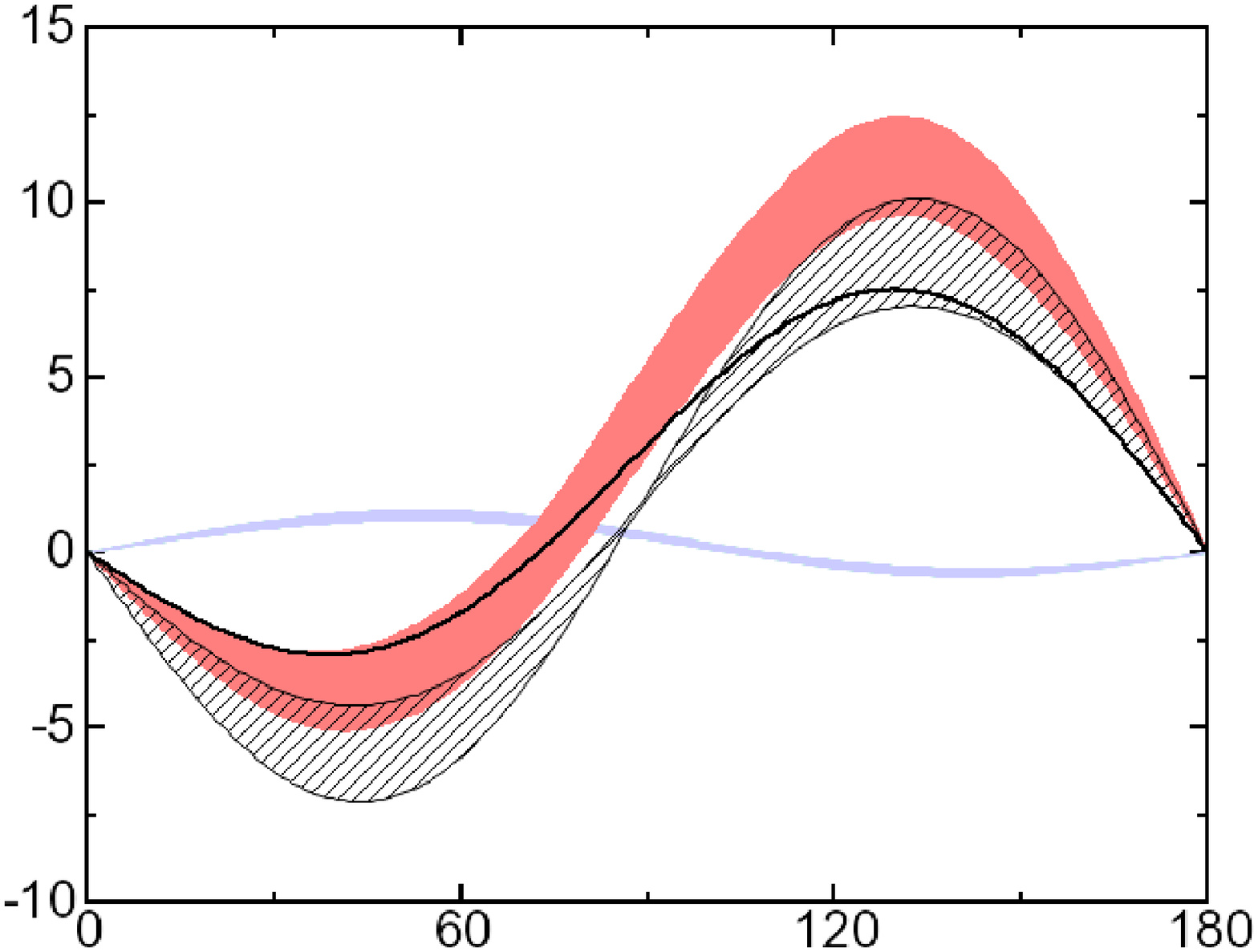}
   \hspace{-0.5mm}\includegraphics[width=5.6cm,height=4.3cm]{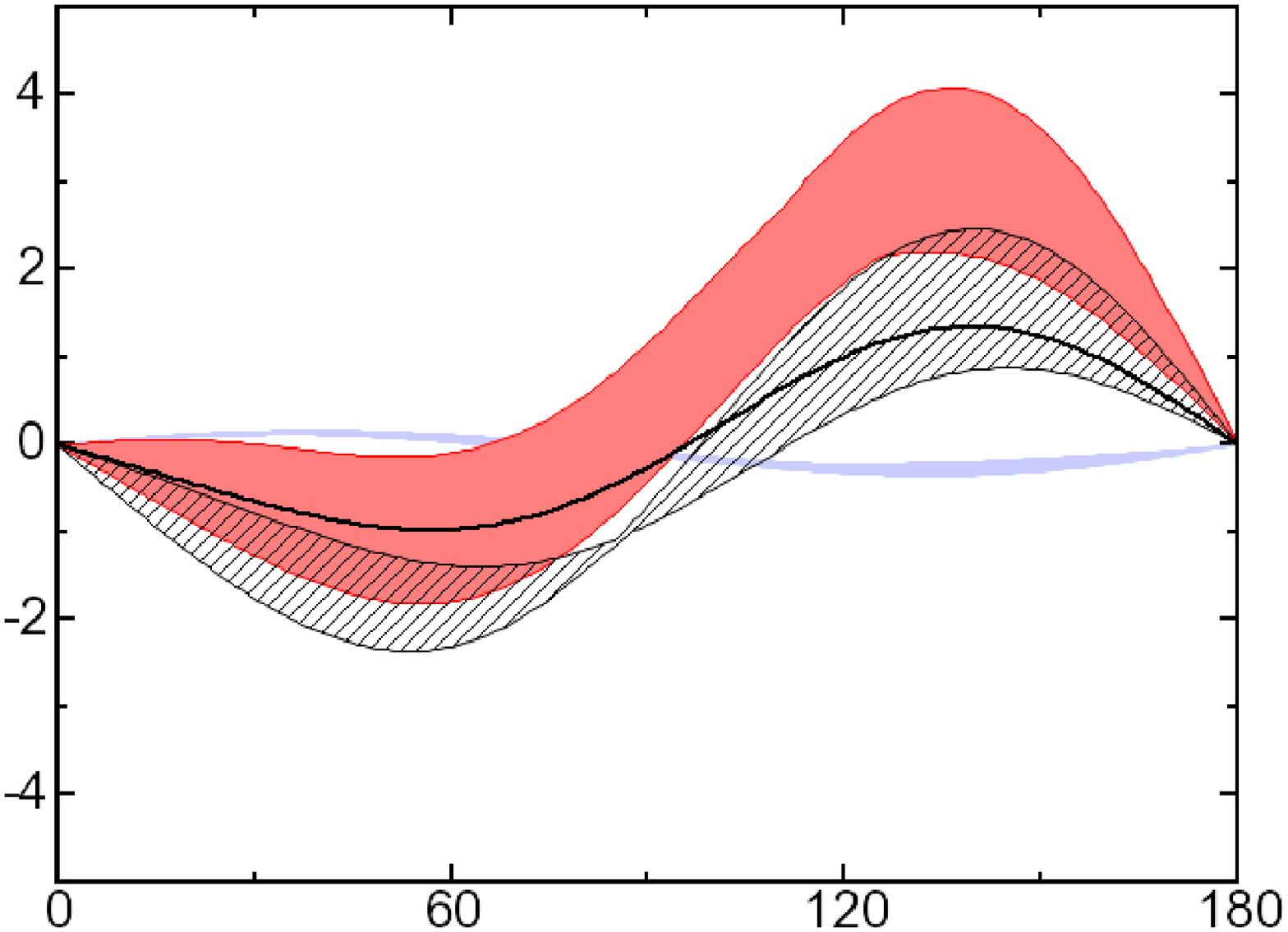}\\
   \hspace{-1mm}\includegraphics[width=6.1cm,height=4.3cm]{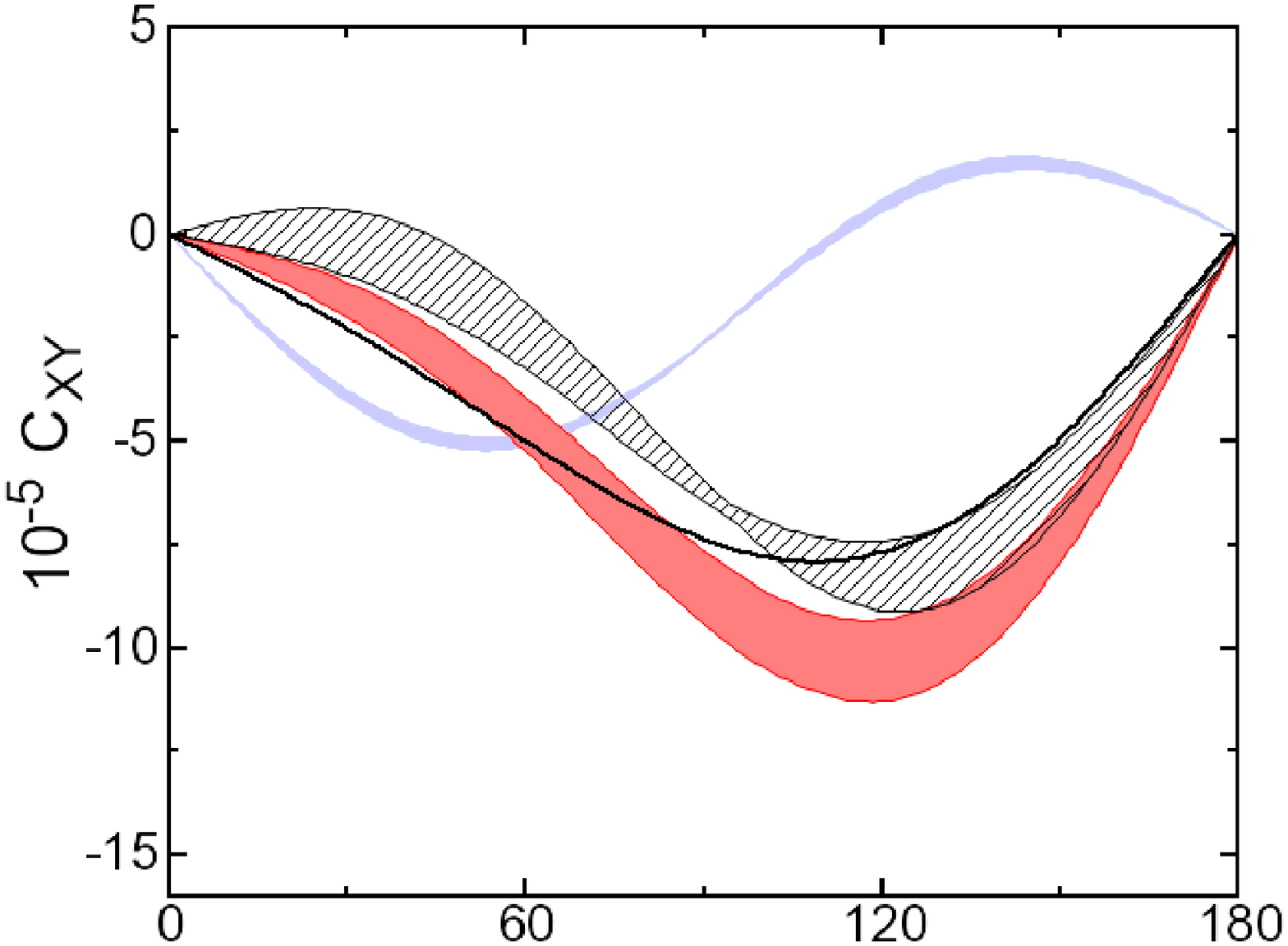}
   \hspace{-0.5mm}\includegraphics[width=5.9cm,height=4.3cm]{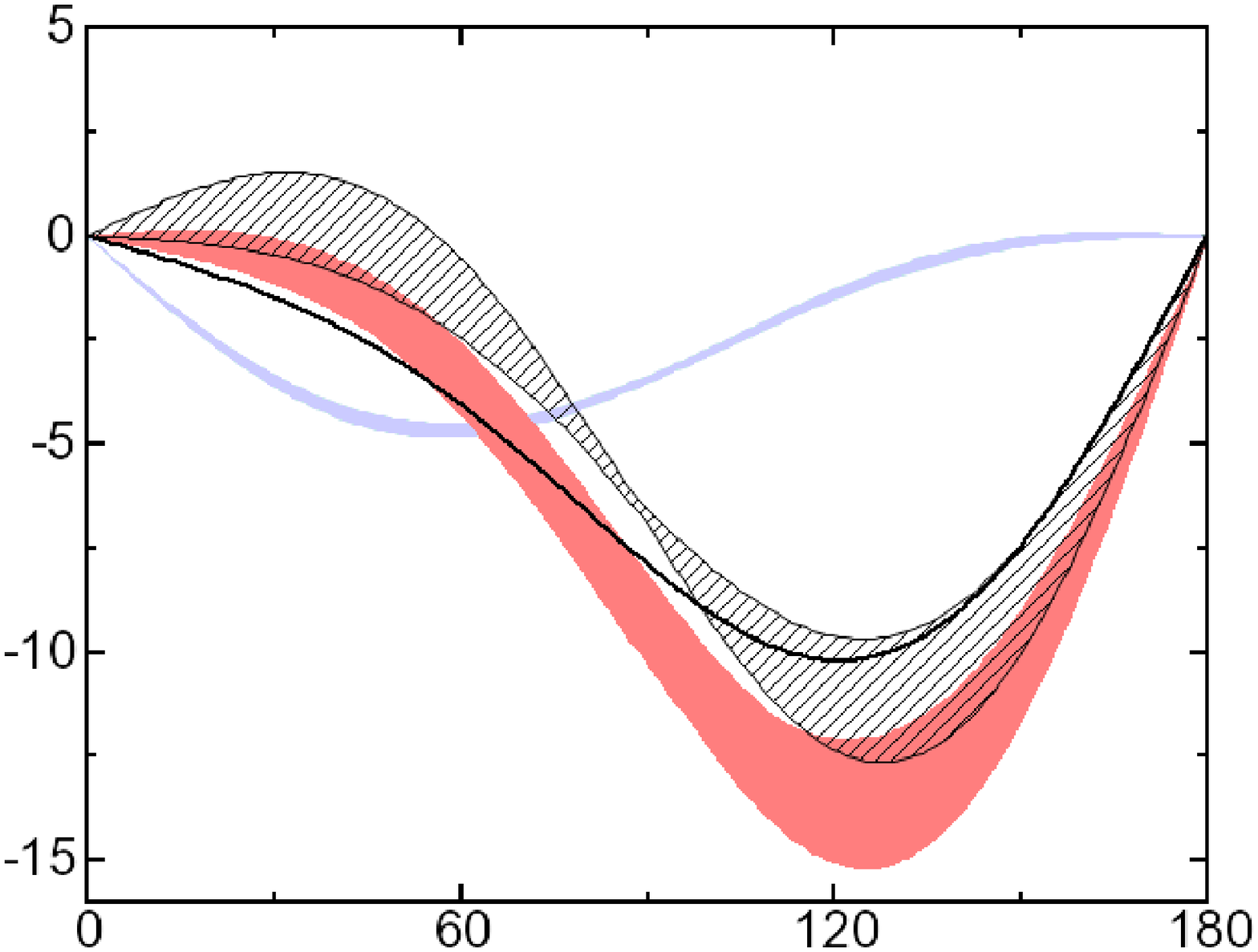}
   \includegraphics[width=5.6cm,height=4.3cm]{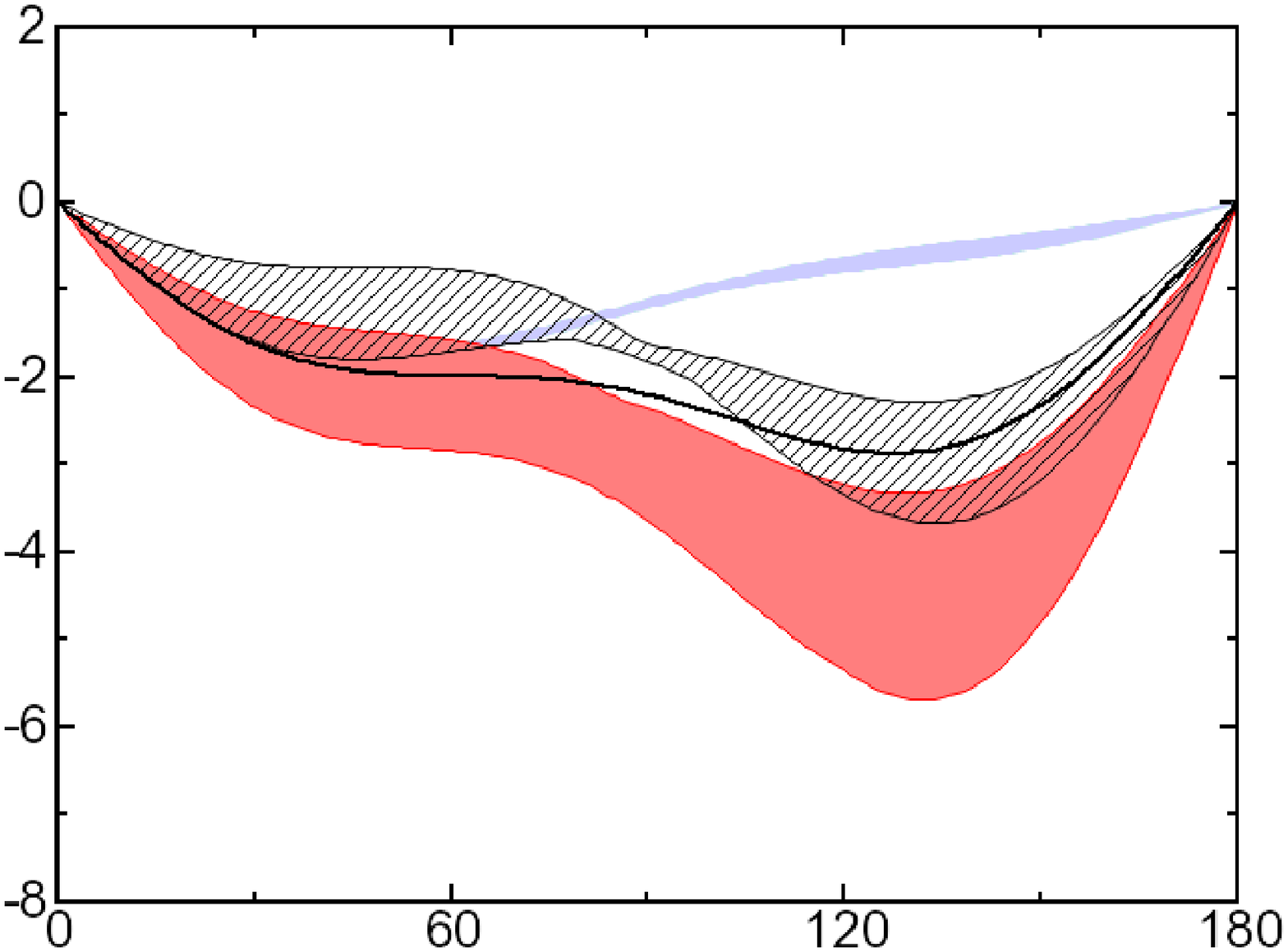}\\
   \hspace{-1mm}\includegraphics[width=6.1cm,height=4.7cm]{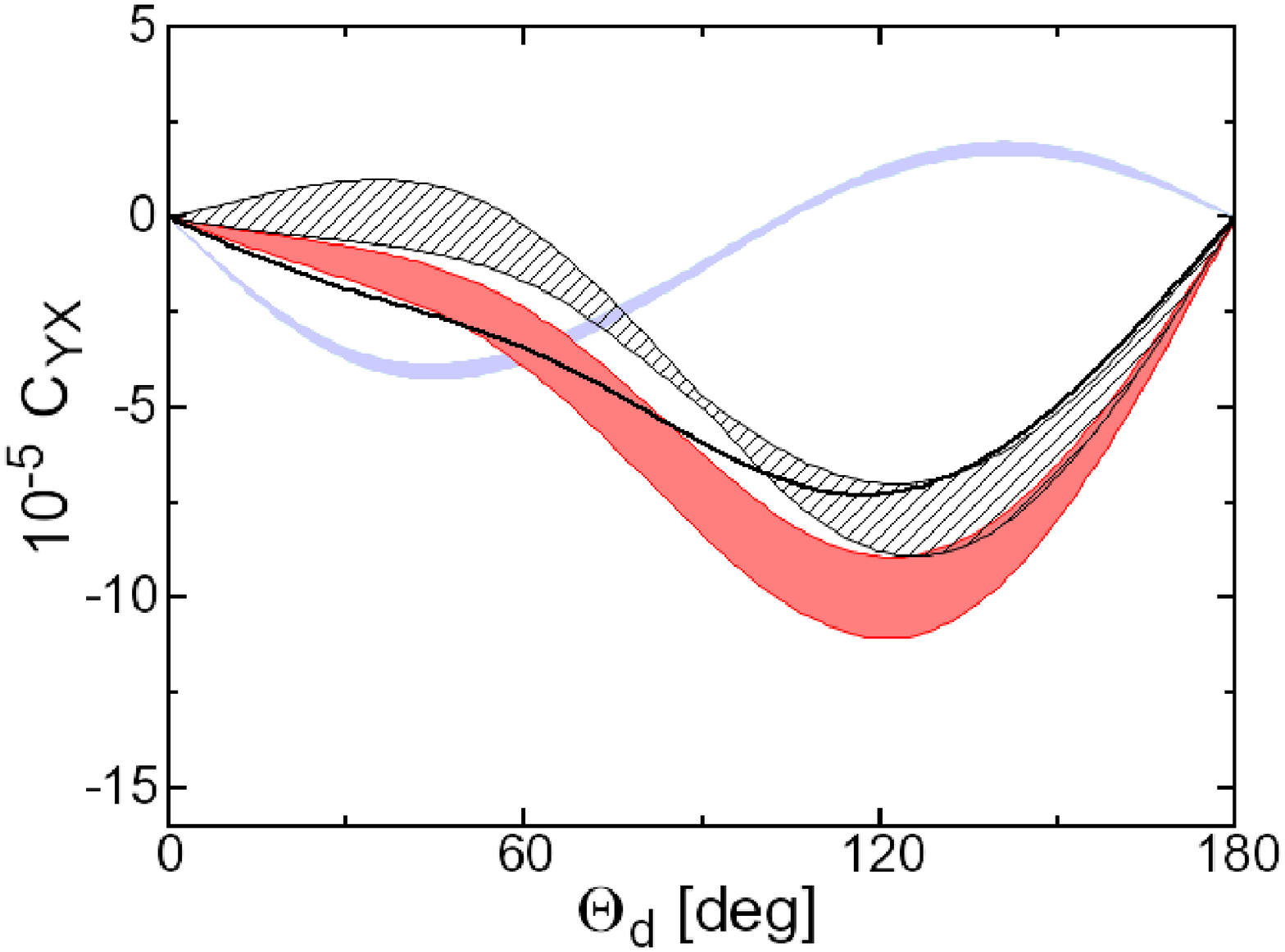}
   \hspace{-0.5mm}\includegraphics[width=5.9cm,height=4.7cm]{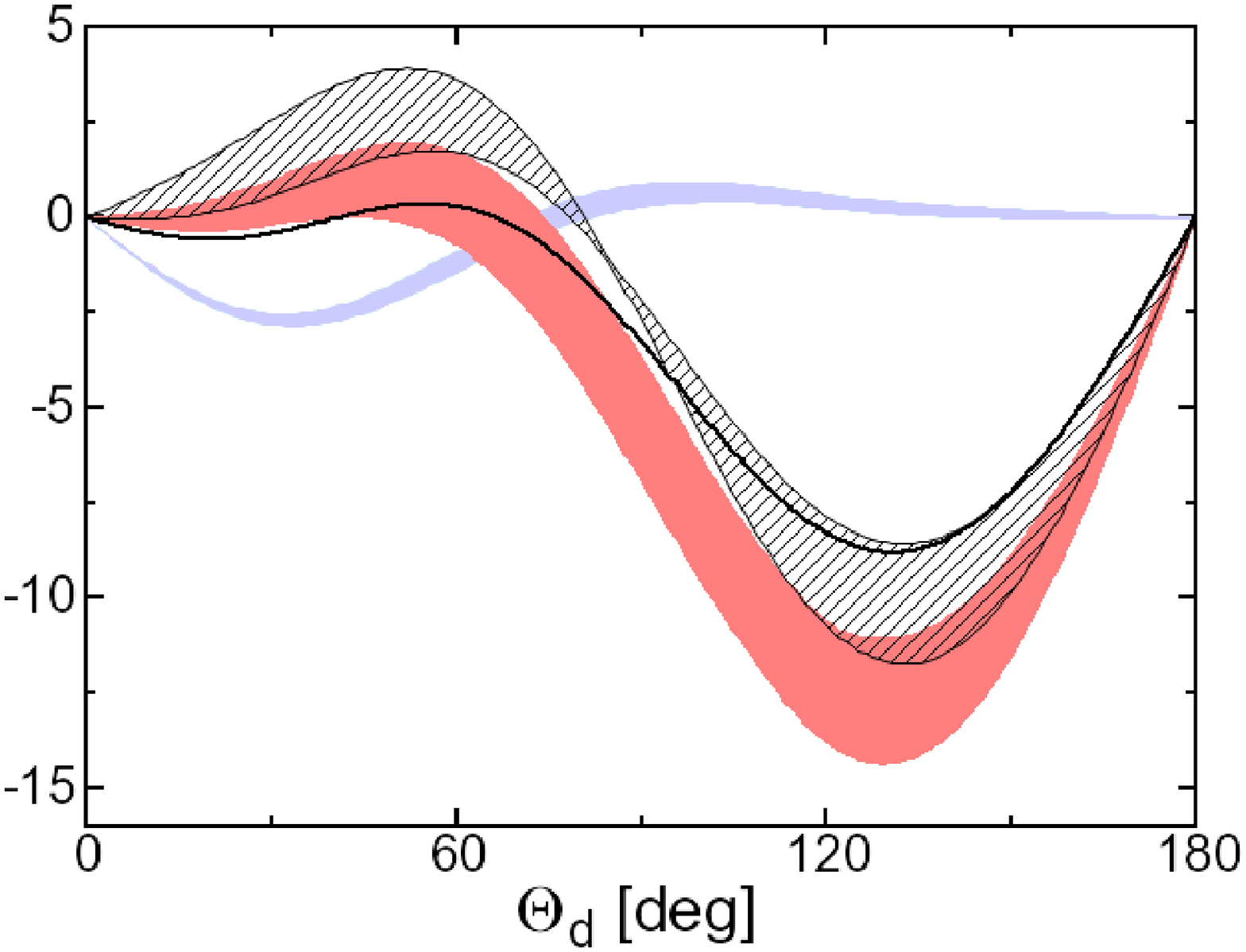}
   \hspace{0mm}\includegraphics[width=5.6cm,height=4.7cm]{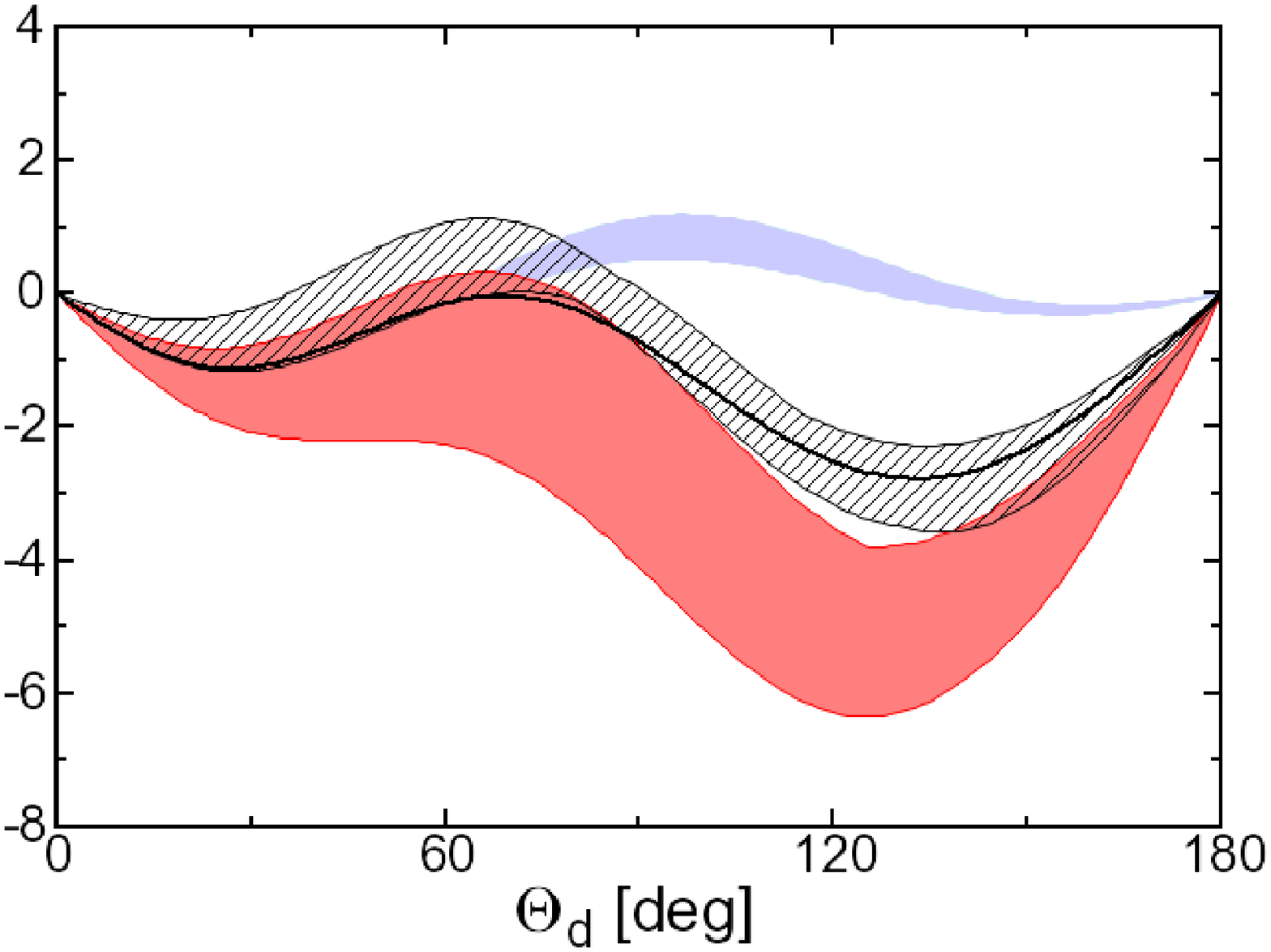}\\
 \end{tabular}
\end{center}
\caption{\label{fg3n2}(color online)
Spin observables for $^3$He two-body photodisintegration
at photon laboratory energy $E_{\gamma}=12$ MeV (left), $E_{\gamma}=20.5$ MeV
(middle) and $E_{\gamma}=50$ MeV (right). 
The upper rows show the analyzing powers for photon ($A_x(\gamma)$) and $^3$He
($A_y(^3$He)). The lower rows show spin correlation coefficients: $C_{XY}$ and 
$C_{YX}$. The bands and lines have the same meaning as in Fig.~\ref{fg3n1}.}
\end{figure}

Next, the results for a few polarization observables are shown in Fig.~\ref{fg3n2}. 
We consider the photon ($A_x^{\gamma}(\theta_d)$)
and the $^3$He ($A_{y}^{^3 \rm He}(\theta_d)$) analyzing powers as well as the spin
correlation coefficients $C_{x,y}^{\gamma,^3 \rm He}(\theta_d)$ and
$C_{y,x}^{\gamma,^3 \rm He}(\theta_d)$. 
In the case of the photon analyzing power $A_x(\gamma)$, 
the prediction bands for the single nucleon current give higher values than
the other, more complete calculations, but the shape of the bands are always similar. 
The TPE bands are broader than OPE bands and overlap with them.
For the $^3$He analyzing power $A_y(^3$He) and the spin correlation coefficients $C_{XY}$ and
$C_{YX}$, we observe that the results based on chiral EFT generate very broad
prediction bands, especially at the highest energy considered. Interestingly,
the results based on the single-nucleon current for these observables are
completely different from the ones involving the MEC. This suggests that these
observables are very sensitive to the details of the meson exchange currents,  
and their proper description will require the inclusion of the subleading OPE
and short-range contributions not considered in the present work. We further
emphasize that the results based on the AV18 potential and the corresponding
MEC agree with the (present) chiral EFT calculation. 

\begin{figure}[t]
\begin{center}
\begin{tabular}{lr}
\hspace{-2.2mm}\includegraphics[width=6.1cm,height=5.1cm]{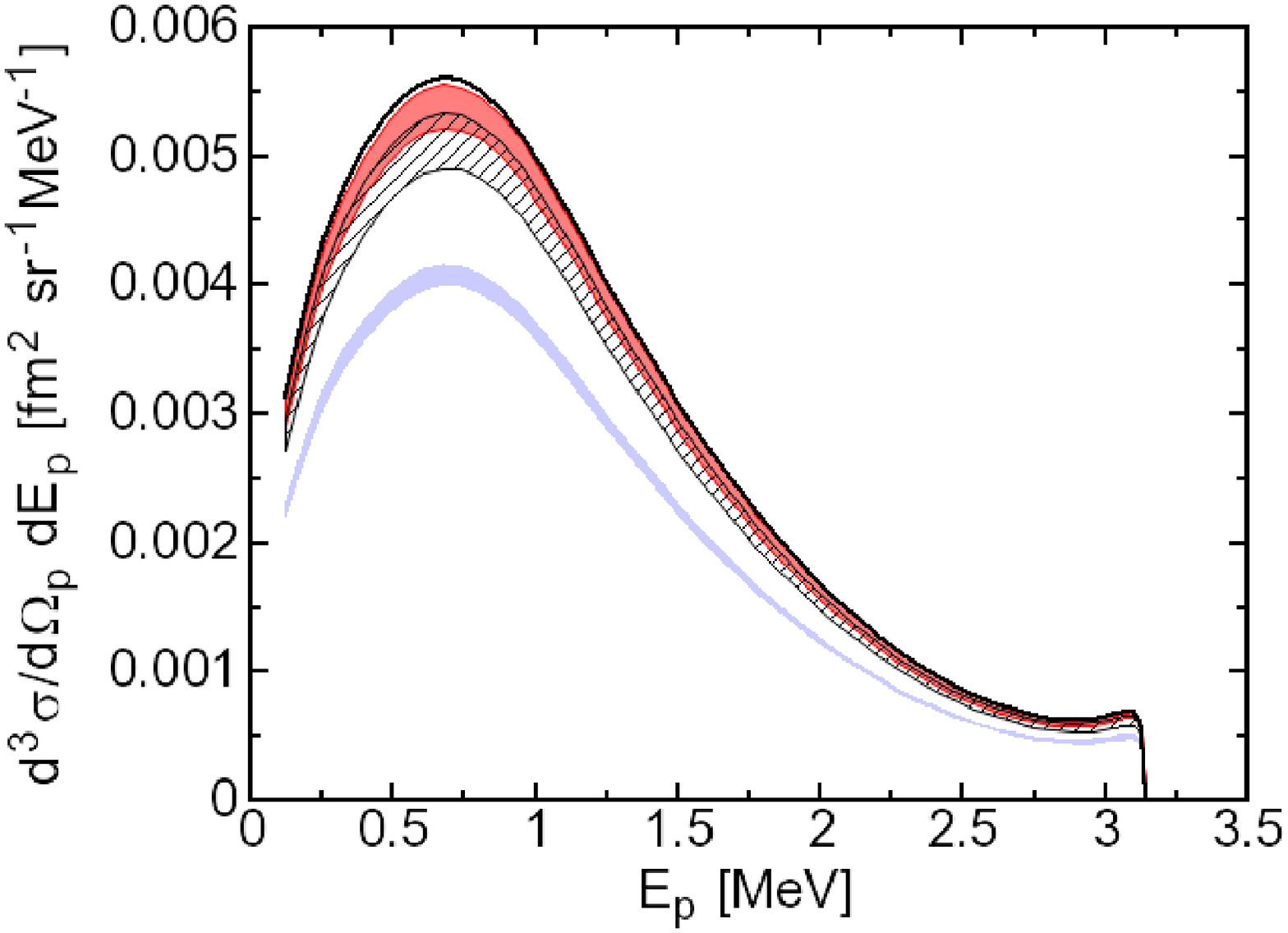}
\hspace{-1.2mm}\includegraphics[width=5.8cm,height=5.1cm]{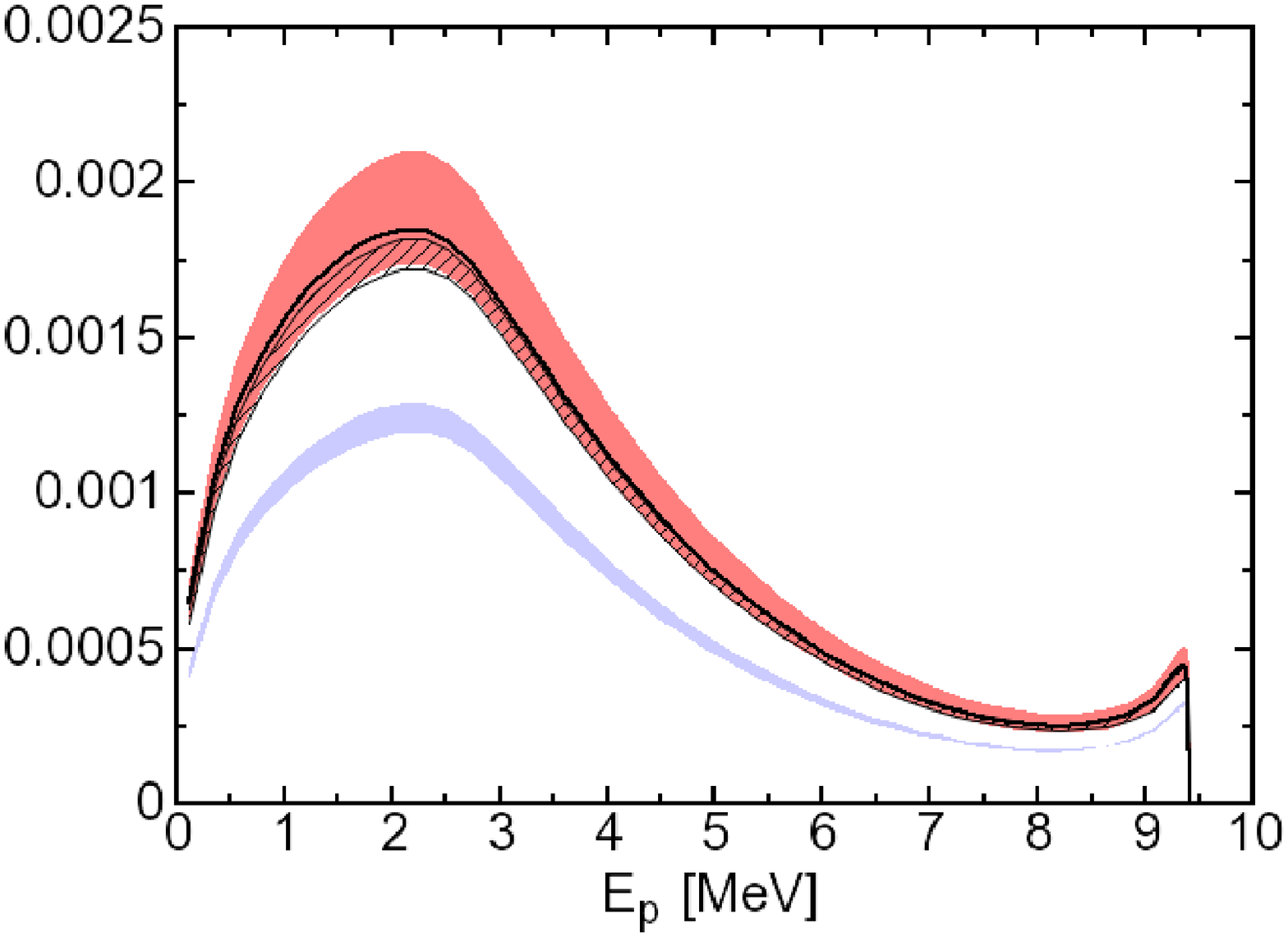}
\hspace{-1.4mm}\includegraphics[width=5.9cm,height=5.1cm]{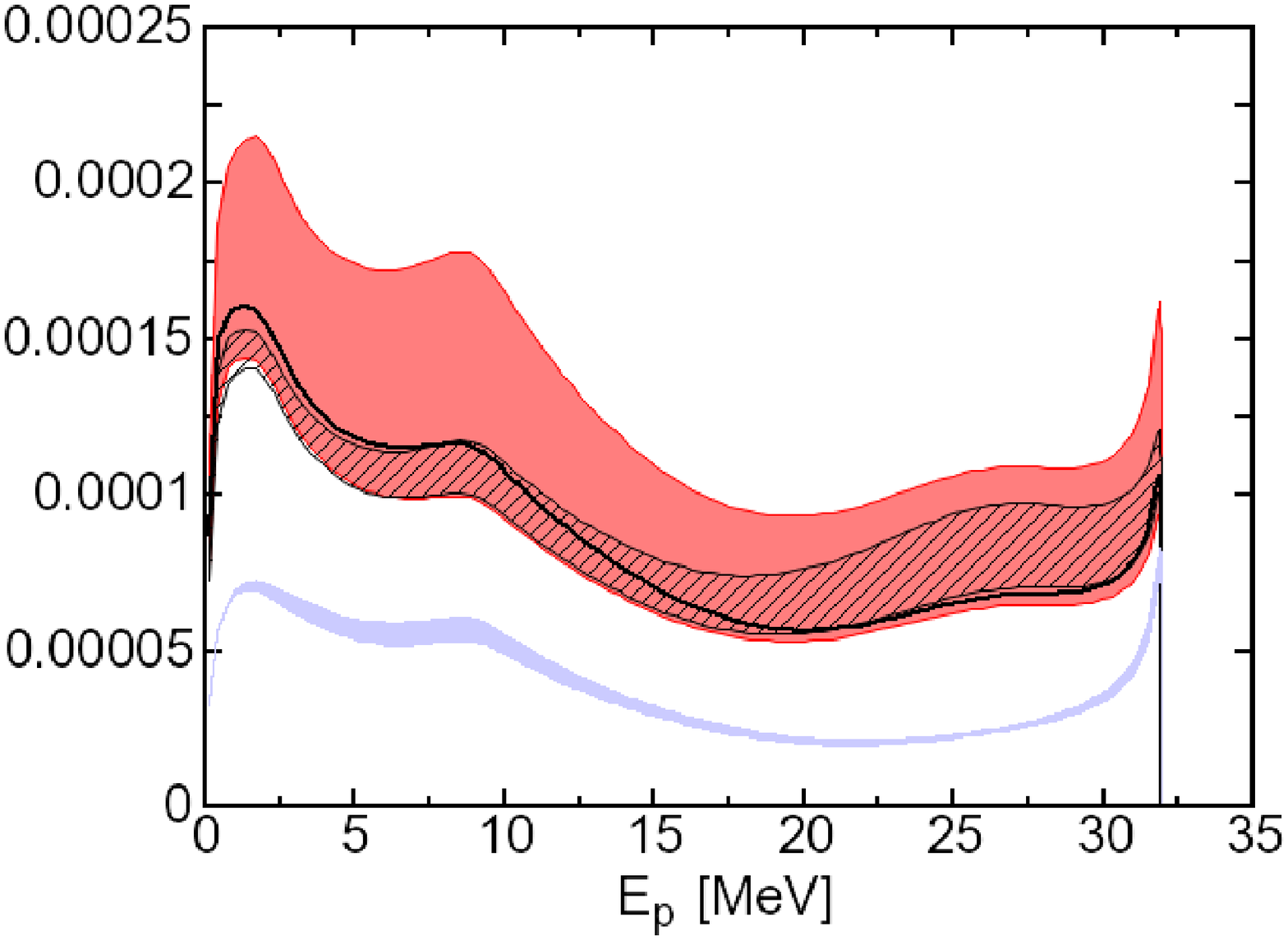}
\end{tabular}
\end{center}
\caption{\label{fg3n3}
Differential cross section for semiexclusive $^3$He three-body photodisintegration
$^3$He$(\gamma,p)pn$ for proton emissions at $\theta=15^o$ and photon laboratory energy $E_{\gamma}=12$ MeV (left), 
$E_{\gamma}=20.5$ MeV (middle) and $E_{\gamma}=50$ MeV (right). 
The bands and lines have the same meaning as in Fig.~\ref{fg3n1}.}
\end{figure}

For the three-body breakup of $^3$He, we only show the semiexclusive
differential cross section $d^3 \sigma/d \Omega_p d E_p$  
(where only one proton would be detected at 15 degrees with respect to the photon beam) at three photon laboratory energies 
$E_{\gamma}$= $12$, $20.5$ and $50$ MeV. 
The calculated cross section is shown as a function of the proton
energies in Fig.~\ref{fg3n3}.
For the lower photon energy (left panel), the obtained bands appear to be relatively narrow,
especially for higher proton energies, where they both coincide with the AV18 results. 
For the lower proton energy the bands become broader. The TPE contributions 
bring the results close to the one obtained within the conventional framework.  
The situation is quite different for the highest photon energy (right panel).
The shape  of the calculated cross section is  much more complicated in this
case. Further, the band resulting from the TPE parts of the current operator
appears to be very broad in the whole range 
of the proton energies.
It remains to be seen whether the inclusion of the missing meson exchange
current contributions will allow to reduce the theoretical uncertainty for the
cross section. 

\section{Conclusions}
In this work we explored the effects of the TPE currents derived recently in
the framework of ChEFT~\cite{TPE:2009} 
in the deuteron and $^3$He photodisintegration reactions. We studied the
role of various ingredients of the chiral 2N current operator in the unpolarized
cross section and several polarization observables. 
As a main outcome of our study, we found that the new
terms in the exchange current operator beyond the well-known one-pion
exchange contribution play an important role for nearly all considered
reactions. In particular, the differential cross section and the photon
analyzing power in the deuteron photodisintegration process and the spin
observables in $^3$He two-body photodisintegration are found to provide
excellent testing ground for probing the fine details of the exchange current
operator. 

We also found that 
the inclusion of the TPE contribution to the current operator alone 
typically results in  very broad bands for the considered observables. 
This behavior is not unexpected. 
The OPE and TPE MEC are computed in the framework of
chiral EFT within the low-momentum expansion and thus feature singular
behavior at short distances (or large momenta). This leads to the observed large sensitivity of the
calculated nuclear matrix elements to the short-distance behavior of the
corresponding wave functions which is strongly scheme and cut-off
dependent. In a \emph{complete} calculation, the cut-off dependence of the
low-energy observables is expected to be strongly reduced by the "running" of
the corresponding short-range current operators, see e.g.~Ref.~\cite{Song:2007bj} 
for the explicit examples of such a behavior in the case of the M1 properties of light
nuclei within the hybrid approach and 
Ref.~\cite{Lepage:1997cs} for an extensive discussion on the (meaning of) renormalization 
in the context of nuclear EFT with a finite cut-off.  
Thus, the strong cut-off dependence in  the obtained incomplete results which
do not include the short-range contributions to the current operator should
not be surprising. We expect that a complete NLO calculation including the 
short-range contact and the subleading OPE contributions to the 2N current operator 
will yield much narrower bands allowing for a quantitative description of 
electromagnetic reactions in a wider kinematical range. 
Work along these lines is in progress. 

\section*{Acknowledgments}
This work was supported in part by the Polish Ministry of Science and Higher
Education (grants N N202 104536, N N202 077435), the Helmholtz Association 
(contract number VH-NG-222) and the European Research Council (ERC-2010-StG 259218 NuclearEFT). 
The numerical calculations have been performed on the supercomputer cluster of the JSC, J\"ulich, Germany.


\vspace{-8mm}
\begin{thebibliography}{99}
\vspace{-1mm}
\bibitem{park:1993}T.~S.~Park, D.-P. Min, and M. Rho, Phys. Rept. \textbf{233}, 341 (1993).
\bibitem{park:2000}T.~S.~Park, K. Kubodera, D. P. Min and M. Rho, Phys. Lett. \textbf{B472}, 232 (2000).
\bibitem{Walzl:2001vb}M.~Walzl and U.-G.~Mei{\ss}ner, Phys. Lett. \textbf{B513}, 37 (2001).
\bibitem{Philips:2009}D.~Phillips, PoS CD\textbf{09}, 066 (2009).
\bibitem{romek1}R.~Skibi{\'n}ski, J.~Golak, H.~Wita{\l}a, W.~Gl\"ockle, A.~Nogga, E.~Epelbaum, Acta Phys. Polon. {\bf B37}, 2905 (2006).
\bibitem{shukla:2009}D.~Shukla, A. Nogga and D. R. Phillips, Nucl. Phys. \textbf{A819}, 98 (2009).
\bibitem{arenhovel:1991} H.~Arenh\"ovel and M. Sanzone, Few-Body Sys. Suppl. \textbf{3}, 1 (1991).
\bibitem{gilman:2002}R.~Gilman and F.~Gross, J. Phys. G, Nucl. Part. Phys. \textbf{28}, R37 (2002).
\bibitem{carlson:1998}J.~Carlson, R. Schiavilla, Rev. Mod. Phys. \textbf{70}, 743 (1998).
\bibitem{marccuci:2009}L.~E.~Marcucci, L. E. Marcucci, A. Kievsky, L. Girlanda, S. Rosati, M. Viviani, Phys. Rev. C \textbf{80}, 034003 (2009).
\bibitem{golak:2005}J.~Golak, R.~Skibi{\'n}ski, H.~Wita{\l}a {\em et al.}, Phys. Rept. \textbf{415}, 89 (2005) and references therein.
\bibitem{evgeny:2006}E.~Epelbaum, Prog. Part. Nucl. Phys. \textbf{57}, 654 (2006).
\bibitem{Epelbaum:2008ga}E.~Epelbaum, H.-W.~Hammer, U.-G.~Mei{\ss}ner, Rev.\ Mod.\ Phys.\  \textbf{81}, 1773 (2009).
\bibitem{pastore:2008}S.~Pastore, R. Schiavilla, and J.L. Goity, Phys. Rev. C \textbf{78}, 064002 (2008).
\bibitem{Pastore:2009is}S.~Pastore, L.~Girlanda, R.~Schiavilla, M.~Viviani, R.~B.~Wiringa, Phys.\ Rev.\ C \textbf{80}, 034004 (2009).
\bibitem{TPE:2009}S.~K\"olling, E. Epelbaum, H. Krebs, U.-G. Mei{\ss}ner, Phys. Rev. C \textbf{80}, 045502 (2009).
\bibitem{ope} V.~V.~Kotlyar, Few Body Syst. \textbf{28}, 35 (2000). 
\bibitem{foldy:1979}L. L. Foldy and J.~A.~Lock, in {\em Mesons In Nuclei}, edited by M.Rho and D. Wilkinson (North-Holland, Amsterdam, 1979), Vol. II.
\bibitem{gloekle:1983}W.~Gl\"{o}ckle, {\em The Quantum Mechanical Few-Body Problem} (Springer-Verlag, Berlin/Heidelberg, 1983).
\bibitem{golak:2010}J.~Golak, D.~Rozp\c{e}dzik, R.~Skibi{\'n}ski {\em et al.}, Eur. Phys. J. A \textbf{43}, 241 (2010).
\bibitem{wiringa:1995} R.~B.~Wiringa, V.~G.~J.~Stoks, R.~Schiavilla, Phys. Rev. C \textbf{51}, 38 (1995).
\bibitem{Riska1}D.~O.~Riska, Phys. Scr. \textbf{31}, 107 (1985).
\bibitem{Riska2}D.~O.~Riska, Phys. Scr. \textbf{31}, 471 (1985).
\bibitem{rachek:2007}I.~A.~Rachek {\em et al.}, Phys. Rev. Lett. \textbf{98}, 182303 (2007).
\bibitem{ying}S.~Ying, E. M. Henley and G. A. Miller, Phys. Rev. C \textbf{38}, 1584 (1988).
\bibitem{naszbs2}A. Nogga, A. Kievsky, H. Kamada, W. Gl\"ockle, L. E. Marcucci, S. Rosati, M. Viviani, Phys. Rev. C \textbf{67} 034004 (2003).
\bibitem{Song:2007bj}Y.~-H.~Song, R.~Lazauskas, T.~-S.~Park, D.~-P.~Min, Phys. Lett. \textbf{B656}, 174 (2007).
\bibitem{Lepage:1997cs}G.~P.~Lepage, arXiv:nucl-th/9706029.
\end{thebibliography}
\end{document}